\documentclass[11pt]{article}
\usepackage[margin=1in]{geometry}
\usepackage{amsmath,amssymb,amsthm,booktabs,array}
\usepackage{graphicx}
\usepackage[round]{natbib}
\usepackage{setspace}
\usepackage{placeins}
\usepackage{xurl}
\usepackage[hidelinks]{hyperref}
\hypersetup{
  pdftitle={Agreements Have a Clock: Compute Growth and the Verification of AI Restraint},
  pdfauthor={Li Gan}
}
\newtheorem{proposition}{Proposition}
\newtheorem{corollary}{Corollary}
\newtheorem{lemma}{Lemma}

\newcommand{\dvRMonthPct}{0.25}
\newcommand{\dvRiskRace}{75}
\newcommand{\dvXCoop}{0.06}

\newcommand{\dvRiskCoop}{6}

\newcommand{\cnDelta}{10.6}
\newcommand{\cnCopyShare}{81}
\newcommand{\cnClosedCopyFour}{94}
\newcommand{\cnClosedUFour}{15.8}
\newcommand{\cnClosedCopySix}{80}
\newcommand{\cnClosedUSix}{3.9}
\newcommand{\cnClosedCopyEight}{58}
\newcommand{\cnClosedUEight}{1.4}
\newcommand{\cnClosureFloorX}{7.7}

\newcommand{\cnBaseLeader}{80}
\newcommand{\cnBaseBoth}{99}
\newcommand{\cnBaseUS}{75}

\newcommand{\cnCatchUp}{12}

\newcommand{\cnSFloorsBoth}{90}

\newcommand{\cnSPauseBlogLeader}{71}
\newcommand{\cnSPauseBlogBoth}{96}

\newcommand{\cnSCeilingLeader}{97}
\newcommand{\cnSCeilingBoth}{99}
\newcommand{\cnSCeilingUS}{65}

\newcommand{\cnSLagLeader}{87}
\newcommand{\cnSLagBoth}{97}
\newcommand{\cnSLagUS}{68}

\newcommand{\cnSSafetyFloorLeader}{57}
\newcommand{\cnSSafetyFloorBoth}{99}

\newcommand{\cnSPauseBlogIncUS}{53}
\newcommand{\cnSPauseBlogIncC}{18}
\newcommand{\cnSPauseBlogCap}{0.0}
\newcommand{\cnSCeilingIncUS}{49}
\newcommand{\cnSCeilingIncC}{47}
\newcommand{\cnSCeilingCap}{1.5}
\newcommand{\cnSLagIncUS}{51}
\newcommand{\cnSLagIncC}{36}
\newcommand{\cnSLagCap}{3.9}
\newcommand{\cnSSafetyFloorIncUS}{41}
\newcommand{\cnSSafetyFloorIncC}{16}
\newcommand{\cnSSafetyFloorCap}{9.9}
\newcommand{\cnBaseIncUS}{67}
\newcommand{\cnBaseIncC}{13}
\newcommand{\cnBaseCap}{9.9}

\newcommand{\cnMaxPauseCap}{0.0}
\newcommand{\cnBestFloorCap}{4.5}
\newcommand{\cnMinFloorCap}{0.0}

\newcommand{\cnCoopLeader}{72}
\newcommand{\cnCoopBoth}{72}
\newcommand{\cnCoopIncUS}{13}
\newcommand{\cnCoopIncC}{59}
\newcommand{\cnCoopCap}{0.0}

\newcommand{\cnCoopGX}{2.5}

\newcommand{\cnCoopGLeader}{57}
\newcommand{\cnCoopGBoth}{83}
\newcommand{\cnCoopGIncUS}{24}
\newcommand{\cnCoopGIncC}{33}
\newcommand{\cnCoopGCap}{2.9}
\newcommand{\cnMaxPause}{12.0}

\newcommand{\cnMaxPauseIncUS}{19}
\newcommand{\cnMaxPauseIncC}{10}
\newcommand{\cnMaxPauseBoth}{94}
\newcommand{\cnMaxPauseLeader}{29}

\newcommand{\cnMinFloorBoth}{67}
\newcommand{\cnMinFloorLeader}{67}
\newcommand{\cnMinFloorTus}{47}

\newcommand{\cnMinFloorIncUS}{4}
\newcommand{\cnMinFloorIncC}{64}
\newcommand{\cnBestFloor}{5}
\newcommand{\cnBestFloorLeader}{41}

\newcommand{\cnBestFloorIncUS}{11}
\newcommand{\cnBestFloorIncC}{31}
\newcommand{\cnBestFloorBoth}{86}
\newcommand{\cnEcGrowthCatchUp}{13.3}
\newcommand{\cnEcGrowthPause}{13.4}

\newcommand{\cnEcGrowthLeader}{24}

\newcommand{\cnEcGrowthGain}{1.3}
\newcommand{\cnEcGrowthSame}{28}

\newcommand{\cnEcGrowthLonger}{3.7}
\newcommand{\cnEcGrowthTotal}{4.8}
\newcommand{\cnEcLevelCatchUp}{13.5}
\newcommand{\cnEcLevelPause}{13.5}

\newcommand{\cnEcLevelLeader}{24}

\newcommand{\cnEcLevelSame}{28}

\newcommand{\cnEcLevelLonger}{4.2}

\newcommand{\cnEcDistillCatchUp}{14.9}
\newcommand{\cnEcDistillPause}{15.2}

\newcommand{\cnEcDistillLeader}{18}

\newcommand{\cnEcDistillGain}{2.9}
\newcommand{\cnEcDistillSame}{25}
\newcommand{\cnEcDistillDirect}{4.4}
\newcommand{\cnEcDistillLonger}{6.9}
\newcommand{\cnEcDistillTotal}{11.2}

\newcommand{\cnMuTwelveLeader}{27}

\newcommand{\cnUnTwelveLeader}{29}

\newcommand{\cnMuTwentyFourBoth}{79}
\newcommand{\cnMuTwentyFourLeader}{7}

\newcommand{\cnPhiOneBase}{94}

\newcommand{\cnPhiOnePauseBoth}{62}

\newcommand{\cnPhiOneFloorsBoth}{51}

\newcommand{\cnSpillBase}{94}

\newcommand{\cnSpillPauseBoth}{47}

\newcommand{\cnSpillFloorsBoth}{35}

\newcommand{\agC}{2.5}
\newcommand{\agT}{3.2}
\newcommand{\agR}{1.0}
\newcommand{\agS}{0.6}
\newcommand{\agLambda}{0.382992}
\newcommand{\agMonitoringMaximum}{0.05556}
\newcommand{\agCriticalWindow}{6.89}
\newcommand{\agEighteenMonitoring}{0.0539}
\newcommand{\agTwentyFourMonitoring}{0.0667}
\newcommand{\agEighteenNetCooperation}{1.044}
\newcommand{\agTwentyFourNetCooperation}{0.699}
\newcommand{\agPauseLength}{36}
\newcommand{\agUSWindow}{12}
\newcommand{\rrTminusR}{2.2}
\newcommand{\rrCminusR}{1.5}
\newcommand{\rrLambda}{0.383}
\newcommand{\rrWindowZero}{12}
\newcommand{\rrMinZero}{0.0319}

\newcommand{\rrDetectProb}{32}
\newcommand{\rrMeanDetect}{31}

\newcommand{\rrLeadBoundary}{0.85}
\newcommand{\rrContFloor}{0.86}
\newcommand{\rrContFloorShare}{0.0086}
\newcommand{\rrLeadTie}{46.9}
\newcommand{\rrOwnRatio}{0.131}

\newcommand{\rrCatchTwelve}{8.61}
\newcommand{\rrCatchTwentyFour}{5.74}
\newcommand{\rrResumeGain}{1.2}
\newcommand{\rrResumeCost}{0.8}
\newcommand{\rrBreakoutProb}{68}
\newcommand{\rrMutualThirtySix}{2.3}
\newcommand{\rrMutualFortyEight}{1.5}
\newcommand{\rrOvertakeHalfPct}{20}
\newcommand{\rrOvertakeCoop}{13.6}
\newcommand{\ackCritLowZero}{3.45}

\newcommand{\ackLastBaseZero}{18.9}

\newcommand{\ackLastBaseOne}{14.8}

\newcommand{\ackLastBaseTwo}{10.9}

\newcommand{\ackGameRangeEnd}{31.8}

\newcommand{\ackEtaLeaderMax}{47.0}

\newcommand{\vsCutTwelve}{22}

\newcommand{\vsCutTwentyFour}{39}
\newcommand{\vsPaperThirtySix}{2.3}
\newcommand{\vsMatchedThirtySix}{1.6}
\newcommand{\vsCutThirtySix}{32}

\newcommand{\vsPzeroScaled}{0.068}
\newcommand{\vsCScaled}{2.6}

\newcommand{\vsCritScaled}{5.44}
\newcommand{\vsLastScaled}{25.5}

\newcommand{\vsLastPzeroFive}{29.2}
\newcommand{\vsPerPointLow}{2.0}
\newcommand{\vsPerPointHigh}{2.1}
\newcommand{\vsEtaBreakEven}{34.2}
\newcommand{\vsEtaBreakEvenPct}{27}

\newcommand{\tkDeltaEighteen}{10.6}
\newcommand{\tkDeltaThirtySix}{11.2}
\newcommand{\tkCopyShareEighteen}{81}
\newcommand{\tkCopyShareThirtySix}{82}
\newcommand{\tkCatchStartEighteen}{12}
\newcommand{\tkCatchStartThirtySix}{12}
\newcommand{\tkCatchTwentyFourEighteen}{5.74}
\newcommand{\tkCatchTwentyFourThirtySix}{5.80}
\newcommand{\tkLastReviewEighteen}{18.9}
\newcommand{\tkLastReviewThirtySix}{19.1}
\newcommand{\tkLastMatchedEighteen}{25.5}
\newcommand{\tkLastMatchedThirtySix}{31.7}
\newcommand{\tkGameRangeEndEighteen}{31.8}
\newcommand{\tkGameRangeEndThirtySix}{31.7}
\newcommand{\tkLastMatchedScaledEighteen}{25.5}
\newcommand{\tkLastMatchedScaledThirtySix}{27.8}
\newcommand{\tkGameRangeEndScaledEighteen}{31.8}
\newcommand{\tkGameRangeEndScaledThirtySix}{71.4}
\newcommand{\tkGlobalRaceEighteen}{80}
\newcommand{\tkGlobalRaceThirtySix}{87}
\newcommand{\tkMutualTwentyFourEighteen}{6.7}
\newcommand{\tkMutualTwentyFourThirtySix}{26.1}
\newcommand{\tkMutualThirtySixEighteen}{2.3}
\newcommand{\tkMutualThirtySixThirtySix}{10.3}
\newcommand{\tkMaxPauseStartEighteen}{0}
\newcommand{\tkMaxPauseStartThirtySix}{14.8}
\newcommand{\tkMaxPauseEighteen}{12.0}
\newcommand{\tkMaxPauseThirtySix}{11.9}
\newcommand{\tkMaxPauseLeaderEighteen}{29}
\newcommand{\tkMaxPauseLeaderThirtySix}{49}

\newcommand{\tkMatchedPauseThirtySix}{1.5}

\newcommand{\tkMutualFollowerThirtySix}{8.9}

\newcommand{\tkReviewGridEnd}{48}
\newcommand{\tkScaledPause}{72}
\newcommand{\tkPauseStartsEnd}{29}
\newcommand{\tkPauseCap}{120}

\newcommand{\crbGrowthThreeLast}{10.9}

\newcommand{\symCopyBaseStart}{12.0}

\newcommand{\symCopyBaseAtTwentyFour}{5.74}

\newcommand{\symCopyBaseLast}{18.9}

\newcommand{\symCopyBaseShare}{81}

\newcommand{\symCopyThreeQStart}{9.1}

\newcommand{\symCopyThreeQLast}{8.0}

\newcommand{\symCopyThreeQShare}{61}

\newcommand{\symCopyHalfStart}{7.6}

\newcommand{\symCopyHalfLast}{2.5}

\newcommand{\symCopyHalfShare}{41}

\newcommand{\symCopyNoneStart}{6.0}

\newcommand{\symCopyNoneOwn}{0.71}
\newcommand{\symCopyEqualStart}{4.4}

\newcommand{\symLeadTwelveAtTwentyFour}{3.86}

\newcommand{\symLeadTwelveLast}{12.4}

\newcommand{\symLeadEighteenLast}{22.8}

\newcommand{\symStakesFortyFiveCrit}{5.39}
\newcommand{\symStakesFortyFiveLast}{25.7}

\newcommand{\symStakesBaseCrit}{6.89}
\newcommand{\symStakesBaseLast}{18.9}

\newcommand{\symStakesBaseXiC}{0.38}

\newcommand{\symStakesFiftyFiveCrit}{8.94}
\newcommand{\symStakesFiftyFiveLast}{10.8}

\newcommand{\symStakesFiftyFiveXiC}{0.48}

\newcommand{\symStakesSixtyCrit}{11.15}
\newcommand{\symStakesSixtyLast}{2.9}

\newcommand{\symStakesSixtyXiC}{0.57}

\newcommand{\symLeaderFixedLast}{18.9}

\newcommand{\symLeaderTwelveLast}{12.4}

\newcommand{\symLeaderEighteenLast}{18.9}

\newcommand{\symLeadStar}{15.5}
\newcommand{\symCritical}{6.89}
\newcommand{\symFallBase}{48}
\newcommand{\symFallNone}{28}

\newcommand{\symLagNeededEqual}{11.6}
\newcommand{\symLagNeededHalf}{7.3}
\newcommand{\symLagNeededQuarter}{4.9}
\newcommand{\symBaseComputePct}{1.7}
\newcommand{\symMonitorRatio}{2.7}
\newcommand{\mcCoop}{2.3}
\newcommand{\mcRaceZero}{80}

\newcommand{\mcOpening}{6.1}
\newcommand{\mcCloseBase}{10.4}

\newcommand{\mcLeaderFinish}{39.6}

\newcommand{\mcRaceZeroLo}{28}

\newcommand{\mcCloseBaseLo}{0.1}
\newcommand{\mcCloseHalfLo}{5.3}

\title{Agreements Have a Clock:\\
Compute Growth and the Verification of AI Restraint}
\author{Li Gan\thanks{Department of Economics, Texas A\&M University. Email: \url{ganli@tamu.edu}. I thank Eric Gan and seminar participants for helpful discussions. AI software (Anthropic's Claude) was used under the author's direction to derive and check the algebra, write the calibration and simulation code, and edit the draft. All output was checked by the author, who is responsible for the content. An interactive companion page, which runs the paper's model in the browser, is at \url{https://ligan98.github.io/frontier-pacing/en/agreements.html}.}}
\date{November 2026}

\begin{document}
\maketitle

\begin{abstract}
Rival powers have agreed to limit nuclear arms, phase out ozone-depleting chemicals, and cut greenhouse gases, though each agreement was hard to reach. This paper asks what an agreement to pause frontier AI needs. A follower advances through own research and diffusion from the leader. Because diffusion fades as the gap closes, a paused leader stays ahead longer than its raw lead suggests. A solved agreement game adds voluntary entry, costly monitoring, secret advances, and credible responses. Cooperation is supportable when the monitoring needed to deter a breakout costs no more than either country gains from cooperating. As compute grows during a pause, a secret advance reaches the frozen frontier sooner, shrinking the response window: agreements have a clock. In an example with a fixed leader window, cooperation is supportable through month \ackLastBaseZero. Verified matched safety research extends this to month \vsLastScaled, and halving the monitoring cost to month \ackGameRangeEnd.
\end{abstract}

\vspace{1em}
\noindent Keywords: Artificial intelligence, international agreements, arms control, verification, geopolitical competition, existential risk.\\
\noindent JEL Classification: F51, F53, D74, O33.

\newpage

\section{Introduction}

Proposals to pace frontier artificial intelligence face a geopolitical objection: a government that restrains its domestic industry may lose technological leadership to a foreign program \citep{amodei2026,aifutures2026}. One concrete answer is Plan A of the AI Futures Project \citep{larsen2026}. The United States and China agree in 2029 to avoid a race to superintelligence, pause frontier development at the level of top human experts in 2035, and restart in 2040. In the verified pause, each side declares its compute, inspectors count chips and audit data centers, training runs take place only in audited clusters, and research is made transparent. Reciprocal restraint of this kind answers the objection only if each country can act in time when the other secretly advances. This paper asks how long that time is, how it changes during a pause, and when it supports a voluntary agreement. The result is a test that Plan A, and any agreement like it, must pass.

Such agreements are hard, but rivals have reached them before. \citet{schellinghalperin1961} define arms control as cooperation between potential enemies. The United States and the Soviet Union limited strategic arms and weighed how to verify the limits, even though such agreements are rare \citep{startverification1989,coevaynman2020}. The Montreal Protocol phased out ozone-depleting chemicals with nearly every country taking part \citep{barrett2003}. The Paris Agreement brought nearly all countries into pledges to cut emissions \citep{keohanevictor2016}. Each relied on terms that the parties could check. Cooperation among rivals is easier when the shared stake is large and the players are few \citep{axelrodkeohane1985}. A pause between two leading AI powers would have both features. The question is what its terms must do.

In short, restraint by one country leaves it exposed, restraint by both pays each of them, and an agreement holds only while each country still has time to answer a secret advance.

First, unilateral restraint leaves a government exposed to foreign progress. A paused leader stays ahead longer than its raw lead because diffusion fades as the follower closes in (Proposition~\ref{prop:copy}). In the calibration the follower is six months behind and draws \cnCopyShare\% of its initial progress from diffusion, so it needs \rrWindowZero{} months to catch a halted leader. That room is finite. A leader that moves slower than the follower's own research is overtaken unless it finishes first, and even a slowdown that keeps the lead shortens later catch-up time (Proposition~\ref{prop:overtake}).

Domestic pacing rules can also raise global risk. In the calibration with the follower racing, a monthly risk ceiling and an R\&D-model lag, which lower the leader's own catastrophe probability, raise the global one (Table~\ref{tab:policies}). Both slow development late in the takeoff period, when the follower can pull alongside the leader at high capability, whereas compute floors spread delay over the takeoff period. Cutting the follower's own research or diffusion instead lengthens catch-up time (Proposition~\ref{prop:buy}).

Second, mutual restraint creates gains for both countries. Both pause capability advancement while continuing safety research, so neither concedes progress to an advancing rival. Under the game's payoff assumptions, lower catastrophe risk and development costs raise each country's payoff while preserving strategic control shares. These gains can finance verification, but each country still has an incentive to advance secretly.

Third, cooperation is supportable when the monitoring needed to deter a secret advance costs no more than either country gains from cooperating (Proposition~\ref{prop:small_agreement}). After a timely finding, a government restarts development when recovering its strategic position is worth the extra cost and risk, as it is under the payoff ordering assumed in Section~\ref{sec:game} (Proposition~\ref{prop:credible_response}). Mutual racing remains an equilibrium too.

Agreements have a clock. Compute keeps growing during a compliant pause, so a secret advance reaches the frozen frontier sooner even though the capability gap is unchanged. The response window, the time a country has to act on a secret advance, shrinks. A shorter window raises the monitoring needed to deter a secret advance, and monitoring costs bound what both countries will pay, so cooperation can stop being supportable before the capability gap closes (Corollary~\ref{cor:game_window}).
In the example, with fixed national values, monitoring costs and a leader window of \agUSWindow{} months, voluntary cooperation is supportable through month \ackLastBaseZero{} (Figure~\ref{fig:clock}). The deterrence and participation bounds meet at a response window of \agCriticalWindow{} months, the critical window. The follower's catch-up time, which sets the window, falls from \rrWindowZero{} months at the start of the pause to \rrCatchTwentyFour{} months at month 24, so it passes that critical value at month \ackLastBaseZero. Halving the cost of monitoring keeps cooperation supportable through month \ackGameRangeEnd, and doubling it rules cooperation out from the start (Table~\ref{tab:clock}).

The clock runs faster when the two countries are more alike. A follower that keeps pace through its own research, or a rival at the same capability, leaves less time to answer a secret advance. A follower with the leader's compute, still six months behind, catches up in \symCopyEqualStart{} months instead of \symCopyBaseStart, so deterring it takes about \symMonitorRatio{} times as much monitoring. Between countries of equal capability an agreement holds only if a breakout must open a lead at least as long as the critical window to be decisive (Section~\ref{sec:symmetric}).

Cheaper monitoring, a faster response, and verified safety commitments can extend cooperation. Matched safety research raises the value of cooperating and lowers the pull of a breakout (Section~\ref{sec:matched_safety}). For Plan A, whose pause is to last five years, those terms decide whether the clock runs out first.

The paper links the takeoff model of \citet{gan2026insights} to a finite bilateral game. The game solves one review block, the decisions at a public review date and a given physical state. Catch-up time less the implementation delay supplies the follower's response window. The leader's window is assumed; if it also shrinks as compute grows and a leader breakout is decisive after 12 months of full-speed progress, it binds first and brings the date forward (Section~\ref{sec:symmetric}). The results rest on assumed national payoffs, a specified follower path, a stipulated closure rule, and an illustrative calibration.

The paper's new element is the link between the physical response window and the price of deterrence. Racing, verification, and the temporary nature of deterrence each have precedents.

International dangerous-technology races are an established subject. \citet{emeryxu2024} solve safety choices under different information about national capabilities. Their central question is how uncertainty and the decisiveness of the race affect risk. Our follower block instead follows capability accumulation, diffusion, and safety over calendar time. It shows how a domestic rule can change foreign exposure and how keeping a capability lead differs from keeping the response window open.

Verification of restraint has its own literature. \citet{baker2023} draws on nuclear verification for AI treaties and stresses quick detection and updating of rules as technology changes; the deterrence bound below prices that quick detection. \citet{dung2025} discuss cooperation, enforcement, and deterrence, and note that technological advances can make deterrence temporary; the agreement clock gives it a physical measure. 

Models of verification built on non-cooperative games study guaranteed detection, the timeliness of inspection, and deterrence \citep{avenhaus1996}; see also \citet{avenhaus2002}. \citet{coevaynman2020} argue that arms control is rare because the transparency needed to assure compliance can threaten security. The monitoring cost here counts resources, so their security cost of transparency lies outside the game. \citet{barrett2003} argues that environmental treaties work when their terms make participation and compliance self-enforcing; the game below asks the same of a pause agreement. \citet{garfinkel2000} show that conflict can be an equilibrium without misperception or incomplete information, because an advantage won today also pays tomorrow. In the game below, the same pull is a secret advance that pays off later, held in check by monitoring. Compute growth then shortens the time in which a finding can still be answered. 

The growth--risk and regulation literatures examine the broader social trade-offs \citep{jones2024,acemoglu2024,bostrom2026,binder2021}; this paper takes national payoffs as given. Companion papers study single-lab commitment bounds, customer-versus-research compute allocation, and racing among domestic labs \citep{gan2026insights,gan2026allocation,gan2026racing}.

Section~\ref{sec:model} presents the takeoff model and calibration. Section~\ref{sec:china} studies the limits of unilateral restraint. Section~\ref{sec:agreement} explains the gains from cooperation before deriving the response window, solving the agreement game, asking how the clock changes when the two countries are more alike, showing how verified safety research lengthens the clock, and asking what the scenario's own risks imply. Section~\ref{sec:results} turns to widening the lead and to pacing instruments when the follower races. Section~\ref{sec:scope} states the limits and Section~\ref{sec:conclusion} concludes.

\section{The Takeoff Model}\label{sec:model}

\subsection{The Pieces in Words}

The model follows one AI frontier through a takeoff period. It is the takeoff model of \citet{gan2026insights} and \citet{gan2026allocation}, with three pieces: compute, capability, and safety knowledge. Compute powers research, capability affects its productivity, and safety knowledge reduces the danger of development.

Compute is the main input of a frontier lab. It means the chips, data centers, and electricity that a lab runs. A lab uses compute in three ways. It serves customers, which is called inference. It does capability research, which trains stronger models. And it does safety research, which makes models less likely to cause harm. The three uses draw on the same chips, so compute spent on one use is not available for the others.

Capability research speeds up as models improve, because AI models help with AI research. The takeoff period starts at the Automated Coder, a model that automates machine learning engineering. It ends at top-expert-dominating AI, a model that surpasses human scientists in every domain. Capability is an index that runs from 0 at the first milestone to 1 at the second.

The catastrophe is a catastrophic failure or a loss of control over AI. The model treats it as a hazard: the chance per month that it happens now, given that it has not happened yet. Hazards add up over time. A hazard of 1 percent a month for a year gives a cumulative hazard of 0.12, or a probability of catastrophe of about 11 percent. Danger has two sources. State hazard is the danger of having a model of a given capability in use, and it accrues per month. Step hazard is the danger of training a stronger model whose properties nobody yet knows, and it accrues per unit of capability gained.

Safety research builds a stock of knowledge, and knowledge lowers both hazards at every later capability level. Knowledge learned now keeps protecting later, stronger models. So slowing down has a benefit and a cost. Each capability level is reached with more knowledge, but more months are spent at each level. Which effect is larger decides whether delay saves risk.

The three pieces interact through time. At full speed a lab puts half of its compute into capability research and 5 percent into safety, and the takeoff period takes 18 months \citep{aifutures2026}. Pacing means moving slower than this, by cutting the research share or by pausing. Installed compute keeps growing whatever the lab does. So a lab that slows down or pauses reaches each later capability level with more compute. A pause that stops capability research still lets safety research continue. Later sections add rivals to this single frontier.

\subsection{Compute and Capability}

Installed compute grows exponentially. Time $t \ge 0$ is measured in months, and total installed computing capacity $K(t)$ grows at rate $g_K$:
\begin{equation}
K(t) = K_0\,e^{g_K t}.
\end{equation}
The calibration sets $g_K=\ln2/6$, so compute doubles every six months. In each period, compute is divided among three uses: customer inference services $i(t)$, capability research $x(t)$, and safety research $s(t)$, satisfying the resource constraint
\begin{equation}
i(t) + x(t) + s(t) = 1, \qquad i(t), x(t), s(t) \ge 0.
\end{equation}

Frontier capability is a state $F(t) \in [0, 1]$. At $F=0$, systems automate machine learning engineering (the Automated Coder milestone). At $F=1$, they surpass human scientific capabilities across all domains (top-expert-dominating AI, or TED-AI). Capability advances according to
\begin{equation}
\dot F(t) = \phi\bigl(F(t)\bigr)\,\bigl(x(t) K(t)\bigr)^\varepsilon,
\label{eq:capability_ode}
\end{equation}
where $\varepsilon \in (0, 1]$ is the research elasticity and $\phi(F)>0$ is the research productivity of a model of capability $F$: more capable models improve the research process. So the pace of capability is research productivity times research capacity, the compute put into research raised to the research elasticity. The results hold for any continuous positive $\phi$. The calibration sets $\phi(F)=\theta e^{\rho F}$, so research at TED-AI is ten times as fast as at the Automated Coder for the same compute. At a fixed share, research capacity $(xK)^\varepsilon$ grows at the rate $\gamma=\varepsilon g_K$. With $\varepsilon=1/2$, quadrupling research compute only doubles the pace at a given capability. So research capacity doubles every year, half the growth rate of compute: $\gamma=\ln2/12$.

Progress accelerates during the full-speed takeoff period. Along an unconstrained full-speed path with baseline research share $x_0 = 0.50$ and safety share $s_0 = 0.05$, the takeoff period from $F=0$ to $F=1$ takes $T_0 = 18$ months. It reaches the halfway point, $F=1/2$, after about 15 months, so the second half of the capability index takes about 3 months.

\subsection{Safety Knowledge and Hazard}

Safety knowledge accumulates through dedicated safety research:
\begin{equation}
\dot S(t) = \sigma\bigl(F(t)\bigr)\,y\bigl(s(t)\bigr),
\label{eq:safety_ode}
\end{equation}
where $\sigma(F)$ is the safety yield, the knowledge a month of safety research adds at the full-speed share $s_0$. The yield is a function of capability, because stronger models change how productive safety research is. The learning multiplier $y(s)$ is continuous and increasing, with $y(s_0)=1$. The calibration sets $y(s)=\psi+(1-\psi)(s/s_0)^\alpha$ with $\psi=\alpha=1/2$: half of baseline learning needs no dedicated compute, and the other half has square-root returns in the safety share. Quadrupling the safety share doubles the second half, so learning runs 1.5 times as fast. With constant $\sigma$, safety knowledge accumulates at a constant calendar-time rate for a fixed safety share. In the calibration $\sigma$ is constant at 0.086 per month, so a year at the full-speed share adds about 1.03 units of knowledge.

At each moment the frontier generates an instantaneous hazard of catastrophic failure or loss of control, $\lambda(t)$. It depends on the capability level, the pace of capability advance, and accumulated safety knowledge:
\begin{equation}
\lambda(t) = \Bigl[ h\bigl(F(t)\bigr) + k\bigl(F(t)\bigr)\,\dot F(t) \Bigr]\,e^{-S(t)},
\label{eq:hazard}
\end{equation}
where $h(F)$ represents state hazard (the hazard of maintaining systems at capability $F$) and $k(F)\dot F$ represents step hazard (the hazard incurred from destabilizing transitions during rapid capability jumps). Knowledge scales both terms by $e^{-S}$. The year of safety research above multiplies the hazard at every later level by $e^{-1.03}$, about 0.36. During a pause, step hazard stops, while state hazard continues at the frozen capability and falls as knowledge grows. Cumulative hazard over the takeoff period is
\begin{equation}
H = \int_0^T \lambda(t)\,dt,
\end{equation}
and the cumulative probability of catastrophe during the takeoff period is $P = 1 - e^{-H}$. We call $H$ the cumulative hazard and $P$ the catastrophe probability. In the baseline calibration, unpaced development carries a catastrophe probability of $P=\dvRiskRace\%$, with state and step hazard contributing equally. That is a cumulative hazard of $\ln4$, about 1.39. The calibration puts three quarters of this hazard in the second half of the capability index, the last 3 months of a full-speed takeoff period.

\subsection{Baseline Numbers}

Table~\ref{tab:calibration} collects the illustrative baseline used below. The numbers describe a scenario; they are not estimates of national capabilities or catastrophe risk. Three of the targets come from the worked example of the AI Futures Project \citeyearpar{aifutures2026}. At full speed the takeoff period takes 18 months and carries a 75 percent probability of catastrophe. A safety-case regime, in which third-party assessors estimate each company's risk and the government caps it, cuts that probability to about 40 percent. We treat 75 percent as the upper end of a range and also report 25 percent, the lower case of \citet{gan2026racing} (Section~\ref{sec:own_risks}). The follower's lag and catch-up targets follow the same source (Section~\ref{sec:copying}). Appendix~\ref{app:calibration} repeats the main results with a slower, 36-month takeoff period (Table~\ref{tab:takeoff36}).

\begin{table}[htbp]
\centering\small
\caption{Scenario inputs, fitted parameters, and model implications}\label{tab:calibration}
\begin{tabular}{@{}p{.23\textwidth}p{.27\textwidth}>{\raggedright\arraybackslash}p{.44\textwidth}@{}}
\toprule
Parameter & Value & Calibration status\\
\midrule
$T_0$; $P_{\rm full}$ & 18 months; 0.75 & Scenario targets from \citet{aifutures2026}\\
$L_{\rm cal}$; $P_{\rm reg}$ & 12 months; 0.40 & Safety-case target from \citet{aifutures2026}, imposed on a stopped-clock pause\\
$\omega$; $q$ & 0.50; 0.75 & Assumed state and late-hazard shares\\
$\varepsilon$; $\rho$ & 0.50; $\ln 10$ & Assumed capability technology, $\phi(F)=\theta e^{\rho F}$\\
$g_K$; $\gamma$ & $\ln2/6$; $\ln2/12$ & Assumed compute growth; implied growth of research capacity\\
$x_0$; $s_0$ & 0.50; 0.05 & Imposed research; assumed safety shares\\
$\psi$; $\alpha$ & 0.50; 0.50 & Assumed learning multiplier $y(s)$\\
$\sigma$ & 0.086116 & Fitted to the 40\% target ($P_{\rm reg}$)\\
$b_0$; $\theta$ & 0.012348; 0.017463 & Fitted to the 18-month takeoff period ($T_0$); implied for $K_0=1$\\
$a_h$; $c_h$ & 0.380750; 2.197225 & Scale fitted to the 75\% target ($P_{\rm full}$); shape set by the assumed $q$\\
Follower lag; $T^*_0$ & 6 months; 12 months & Lag and catch-up targets from \citet{aifutures2026}\\
$\sigma_B/\sigma$ & 0.50 & Assumed follower safety yield\\
$F_B(0)$ & -0.058490 & Model implied by the initial lag\\
$a_B$; $\delta$ & 0.001616; 0.106318 & Jointly fitted to the lag and catch-up targets\\
Initial diffusion share & 81.49\% & Model implied, not directly measured\\
\bottomrule
\end{tabular}
\par\smallskip
\begin{minipage}{\textwidth}\footnotesize
\emph{Notes:} Time is measured in months and $K_0=1$. Targets from \citet{aifutures2026} are judgments in its worked example, used here as illustrative scenario inputs; assumed parameters are not estimated here. The pause target holds the compute clock fixed and retains baseline safety learning. Research capacity grows at $\gamma=\varepsilon g_K$. The fitted hazard density is $a_h e^{c_h F}$ on the full-speed path. The baseline ends both national hazards at the first terminal arrival. Complete values and alternative closure outcomes are recorded in the replication results.
\end{minipage}
\end{table}

\section{The Problem Without an Agreement}\label{sec:china}

Unilateral restraint exposes a country to an advancing rival. Even if domestic labs are coordinated under a common rule \citep{gan2026racing}, a government faces an external constraint: the fear of falling behind an unpaced rival. The rival faces the same fear, and that shared fear is what makes a mutual agreement worth having. The leader and follower are generic roles; their productivity and safety differences are assumptions of the scenario. Diffusion sets catch-up time, and speed keeps the lead.

\subsection{The Follower Frontier and Diffusion Dynamics}\label{sec:china_setup}

The leader frontier $F(t)$ follows the domestic model. The follower operates an independent frontier $F_B(t)$, which starts below the Automated Coder milestone. The six-month calendar-lag calibration gives $F_B(0)=-0.05849$. We extend the progress law below zero until the follower reaches that milestone, and take $\phi$ to be continuous and positive on $[F_B(0),1]$. Safety knowledge accumulates at its baseline rate throughout, but follower catastrophe hazard is zero while $F_B<0$ and becomes active under the shared hazard law at $F_B=0$. So negative capability is a pre-milestone index, not an additional interval of the normalized leader takeoff period.

The follower advances through two distinct mechanisms:
\begin{equation}
\dot F_B(t) = \underbrace{a_B\,e^{\gamma_B t}\,\frac{\phi(F_B(t))}{\phi(0)}}_{\text{own research}} + \underbrace{\delta\,\max\bigl(F(t) - F_B(t),\,0\bigr)}_{\text{algorithmic diffusion}}.
\label{eq:china_progress}
\end{equation}
Own research grows with the follower's research capacity and, through the same research productivity $\phi$ as the leader's, with its capability. The ratio $\phi(F_B)/\phi(0)$ measures that productivity relative to the Automated Coder milestone, so $a_B$ is the follower's speed at date zero at that milestone. In the calibration the ratio rises tenfold between the two milestones. Here $\gamma_B$ is the growth of effective research capacity after research diminishing returns. In the equal-growth baseline it equals $\gamma=\varepsilon g_K$, half the growth rate of compute in the calibration.

The second term represents algorithmic diffusion: the transfer of knowledge developed at the leading frontier. Diffusion closes a fraction $\delta > 0$ of the capability gap $G(t) \equiv F(t) - F_B(t)$ each month. If the follower overtakes the leader ($F_B > F$), diffusion ceases. The follower starts $t=0$ at a six-month calendar lag behind the leader and operates at full unconstrained speed.

We define four distinct concepts:
\begin{enumerate}
\item \emph{Capability gap ($G(t)$):} The absolute distance $F(t) - F_B(t)$.
\item \emph{Calendar lag:} The elapsed time since the leader was at the follower's current capability.
\item \emph{Catch-up time:} The number of months the leader could immediately halt while remaining ahead. In the constant-speed approximation this is $T^*(G)$; with growing compute it also depends on the calendar date and frontier states.
\item \emph{Keeping the lead:} The condition that $F(t) > F_B(t)$ for all $t$ until the leader reaches $F=1$.
\end{enumerate}

The follower's frontier carries the same state and step hazards \eqref{eq:hazard} as the leader's, mitigated by its own safety knowledge $S_B(t)$. We assume $S_B$ grows at half the leader's full-speed rate: $\dot S_B=\sigma_B$ with $\sigma_B=\sigma/2$, where $\sigma$ is the calibration's constant safety yield. Because the follower's safety practices are unpaced, its frontier represents an independent source of global catastrophe. Section~\ref{sec:matched_safety} asks what changes when an agreement commits the follower to the leader's safety research.

We evaluate two distinct closure rules:
\begin{enumerate}
\item \emph{First-arrival closure (baseline):} The takeoff period ends when the first national frontier reaches its terminal milestone, whichever country arrives first. This assumes that the first arrival terminates both countries' hazards, for example through a defensive intervention. It is a substantive termination assumption, not a result of the model.
\item \emph{Independent closure (both run to completion):} Each frontier generates hazard until its own terminal arrival. The follower continues after a leader arrival until $F_B=1$. In the safety-spillover scenario, leader safety learning continues after its arrival at baseline rate $\sigma$.
\end{enumerate}

\subsection{Catch-Up Time and the Dominance of Diffusion}\label{sec:copying}

Diffusion is fast while the gap is wide and fades as the gap closes, so a halted leader stays ahead longer than the raw lead suggests. To see how, take constant own research speed $a$ and initial gap $G_0$. The derivation takes three steps.

\emph{Step 1: the gap after a halt.} When the leader halts, the gap shrinks at the follower's speed, $a+\delta G$. Own research closes a fixed amount each month; diffusion closes a fixed share of what is left.

\emph{Step 2: the catch-up date.} This linear law makes the gap decay toward $-a/\delta$ at rate $\delta$. It reaches zero after $\ln(1+u)/\delta$ months, where $u=\delta G_0/a$ compares diffusion with own research at the halt.

\emph{Step 3: the raw lead.} A follower keeping pace covers the gap at speed $a+\delta G_0$, so the raw lead is $G_0/(a+\delta G_0)$ months. Dividing the catch-up time by the raw lead gives a function of $u$ alone.

\begin{proposition}[Catch-Up Dynamics and Diffusion Dominance]\label{prop:copy}
Let the follower's own research speed be $a > 0$, and let diffusion close the share $\delta$ of the gap, which is $G_0$ when the leader halts. If the leader halts ($\dot F = 0$), the follower catches up in
\begin{equation}
T^* = \frac{1}{\delta}\ln(1 + u), \qquad u \equiv \frac{\delta G_0}{a},
\label{eq:catchup_formula}
\end{equation}
where $u$ is the ratio of diffusion to own research at the halt. If the follower was keeping pace before the halt ($\dot F_B = \dot F$), its raw lead is $\text{raw} = G_0/(a + \delta G_0)$, and the ratio of catch-up time to raw lead is
\begin{equation}
\frac{T^*}{\text{raw}} = \frac{(1+u)\ln(1+u)}{u}.
\label{eq:ratio_formula}
\end{equation}
\end{proposition}

The proof is in Appendix~\ref{app:proof_copy}.

In words, catch-up takes longer than the raw lead because diffusion fades as the follower approaches a halted leader ($G \to 0$). Only slow own research $a$ then remains. The ratio $f(u) = (1+u)\ln(1+u)/u$ strictly increases from $f(0) = 1$ to $\infty$, so a catch-up time much larger than the raw lead implies a large initial diffusion ratio $u$.

For the generic leader--follower scenario, we use the 4--8-month raw lead and roughly 12-month catch-up time proposed by the AI Futures Project \citeyearpar{aifutures2026}. Inverting equation~\eqref{eq:ratio_formula} in the constant-speed approximation gives the diffusion share. With a 12-month catch-up, a raw lead of 4, 6 or 8 months means $T^*/\text{raw}=3.0$, 2.0 or 1.5, so $u\approx\cnClosedUFour$, \cnClosedUSix{} or \cnClosedUEight. Diffusion then accounts for $u/(1+u)=\cnClosedCopyFour$, \cnClosedCopySix{} or \cnClosedCopyEight{} percent of the follower's progress.
In the accelerating calibration, the follower closes $\delta = \cnDelta\%$ of the capability gap per month, and diffusion accounts for $\cnCopyShare\%$ of its initial progress. This share is model-implied rather than directly measured: the calibration matches a six-month calendar lag, the leader's speed at that earlier date, and the proposed catch-up time. It differs from the constant-speed inversion because own progress and recursive self-improvement change during catch-up.

\subsection{Overtaking and Catch-Up Time}\label{sec:overtake}

A leader that slows instead of halting keeps its lead only if it moves fast enough. Let it advance at a steady pace $\nu > 0$.

\emph{Step 1: the gap's law.} The gap now changes at $\nu-a-\delta G$: the leader adds $\nu$, own research takes away $a$, and diffusion takes away $\delta G$.

\emph{Step 2: the resting point.} The gap moves toward the level $G^*=(\nu-a)/\delta$ at which these forces balance.

\emph{Step 3: overtaking.} If $\nu\ge a$, the resting point is not negative and the gap never reaches zero. If $\nu<a$, it is negative, and the gap crosses zero in finite time. Either way, a shrinking gap shortens catch-up time.

\begin{proposition}[The Overtaking Threshold]\label{prop:overtake}
Let the follower's own research advance at rate $a$, and let the leader advance at constant speed $\nu$ from initial gap $G_0 > 0$.
\begin{enumerate}
\item[(a)] Until the first crossing of zero, the capability gap evolves as $G(t) = G^* + (G_0 - G^*)e^{-\delta t}$, where
\begin{equation}
G^* = \frac{\nu - a}{\delta}.
\label{eq:asymptotic_gap}
\end{equation}
\item[(b)] If $\nu \ge a$, the follower never overtakes ($G(t) > 0$ at every finite date). If $\nu<a$ and both programs continue indefinitely, the follower overtakes in finite time. A finite terminal milestone can occur before that crossing.
\item[(c)] The catch-up time at date $t$ is $T^*(G(t)) = \frac{1}{\delta}\ln(1 + \delta G(t)/a)$. Any speed $\nu < a + \delta G_0$ causes the gap to contract toward $G^*$, steadily eroding the leader's catch-up time even if the follower never overtakes.
\end{enumerate}
\end{proposition}

The proof is in Appendix~\ref{app:proof_overtake}.

Part (b) says the follower cannot overtake a leader that moves at least as fast as the follower's own research. A slower leader is overtaken if both programs continue long enough. Part (c) says the leader's catch-up time shrinks as the gap shrinks, even before any overtaking.

A positive research share alone does not preserve leadership. In the accelerating model, diffusion vanishes at the boundary $F_B=F$. A sufficient no-crossing condition is
\begin{equation}
\phi(F)\,(x(t)K(t))^\varepsilon\ge a_Be^{\gamma_Bt}\,\frac{\phi(F)}{\phi(0)}
\quad\text{whenever }F_B=F.
\label{eq:boundary_speed}
\end{equation}
At the boundary both frontiers sit at the same capability, so research productivity multiplies both sides equally and drops out. Under equal research-capacity growth, $\gamma_B=\gamma$, a constant share satisfies the condition when $x\ge x_0(a_B/b_0)^{1/\varepsilon}\approx\rrContFloor\%$, where $b_0=\phi(0)(x_0K_0)^\varepsilon$ is the leader's initial full-speed pace.\footnote{With locally Lipschitz progress laws this condition makes the nonnegative-gap region invariant, so a strictly positive initial gap stays positive at finite dates.} At the same capability and date, the follower's own research runs at $a_B/b_0\approx\rrOwnRatio$ of the leader's full-speed pace ($b_0$ in Table~\ref{tab:calibration}). With $\varepsilon=1/2$ the floor is $x_0$ times the square of that ratio, $0.5\times\rrOwnRatio^2\approx\rrContFloorShare$. 

This continuing-horizon floor is close to the smallest share that keeps the lead only until a finite terminal milestone, \rrLeadBoundary\%, because at that smaller share the follower draws level just as the leader completes its takeoff period, at month \rrLeadTie. Below the floor the lead can be lost. In the same calibration, a 0.5\% research share with 25\% safety allows overtaking after about \rrOvertakeHalfPct{} months. The diffusion mechanism creates room for restraint but does not remove these constraints.

\FloatBarrier
\section{Gains from Cooperation and the Agreement Clock}\label{sec:agreement}

\subsection{Gains from Mutual Restraint}\label{sec:gains}

Reciprocal restraint can benefit two rivals even when neither is willing to restrain itself alone. The security dilemma is that a country's effort to protect its own position can induce a response that leaves both worse off \citep{jervis1978}.

Under compliance, both countries stop own capability research and capability-enhancing adoption while continuing their safety programs. Their capability gap stays fixed, so both accumulate safety knowledge without conceding progress to an advancing rival.

If compliance holds, longer mutual pauses remove more risk under first-arrival closure. A 12-month mutual pause with safety research lowers the global catastrophe probability to \cnMuTwelveLeader\%, against \cnBaseLeader\% when both countries race at full speed and \cnUnTwelveLeader\% under the corresponding unilateral pause. A unilateral pause cannot run past \cnMaxPause{} months without letting the follower overtake, but a mutual pause can. A 24-month mutual pause gives \cnMuTwentyFourLeader\% under first-arrival closure and \cnMuTwentyFourBoth\% under independent closure, because under independent closure the follower's takeoff period still runs to completion. Under first-arrival closure the probability falls further, to \rrMutualThirtySix\% at 36 months and \rrMutualFortyEight\% at 48. So choosing a duration needs delay costs as well as risk. How much of this risk reduction is achievable depends on how long the response window stays open.

The game asks whether each country's national gain can support an agreement. Under its payoff assumptions, cooperation lowers catastrophe risk and development costs while preserving strategic control shares. These assumed gains differ from the simulated global risk reductions, and Section~\ref{sec:matched_safety} shows how a simulated cut enters the game. They can finance verification, but agreement also needs terms that deter a secret advance.

\subsection{The Response Window}\label{sec:mutual}

Nuclear arms control addressed this problem through agreed limits, monitoring, and response time. A START verification assessment explicitly distinguished collecting information, evaluating the strategic significance of cheating, and having timely countermeasures \citep{startverification1989}. These mechanisms motivate declarations of covered compute facilities, auditable training records, and agreed inspection procedures; their technical effectiveness in AI is an additional implementation question.

The economic question is whether both countries prefer to comply. A monitoring system can reveal a secret advance and still leave the country that advanced better off than under compliance. We distinguish the strategic use of the response window from a threat that would never be carried out, following the bargaining logic of \citet{schelling1966}.

Suppose both countries halt capability advancement while continuing safety research, keeping the capability gap fixed under compliance. If the follower secretly advances, the leader stays paused until its alarm response finishes, after a detection delay $D$ and an implementation delay $\ell$, or until the remaining agreed pause $R$ ends. Until then the gap shrinks along the halt path of Proposition~\ref{prop:copy}. So the leader keeps a positive lead when it restarts before the follower's catch-up time, and keeps it afterwards if it then moves at least as fast as the follower's own research (Lemma~\ref{prop:mutual} in Appendix~\ref{app:proofs}). In the accelerating model, define $T_{\mathrm{catch}}(d;F(d),F_B(d))$ as the first time to reach the frozen leader frontier if the follower secretly advances its baseline program at treaty date $d$. Restarting with a positive gap requires
\begin{equation}
\min\{D+\ell,R(d)\}<T_{\mathrm{catch}}(d;F(d),F_B(d)),
\label{eq:physical_response}
\end{equation}
where the leader's exposure, the shorter of its alarm response time and the remaining pause, must be less than the follower's catch-up time. Maintaining the boundary speed condition \eqref{eq:boundary_speed} thereafter is sufficient to preserve that positive gap.

Compute keeps growing during a compliant pause, so a fixed capability gap does not preserve a fixed catch-up time. The calibrated catch-up time falls from \rrWindowZero{} months at the start to \rrCatchTwelve{} months after 12 months of compliance and to \rrCatchTwentyFour{} months after 24 months. Table~\ref{tab:agreement_game} reports these times alongside the game solution. 

A fixed response standard fails as catch-up time falls. For example, a secret advance at month 12 of a 24-month pause permits catch-up after \rrCatchTwelve{} months. Detection ten months later meets the one-year standard that suits the start, but occurs after overtaking. A longer agreement needs a date-dependent response standard, an earlier scheduled restart, or additional safeguards.

\subsection{A Solved Agreement Game}\label{sec:game}

\subsubsection{Payoffs and Monitoring}

The game covers one review block: the round of decisions at one public review date. Each government chooses whether to pause or advance its capability program. The block ends in a terminal outcome after any timely finding and response. An advancing program is committed until that outcome; a late breakout is absorbing. These restrictions keep the game small, because there are no subsequent choices to conceal again, stop the advancing program, or impose a punishment after breakout.

Index the leader by $A$ and the follower by $B$. Before monitoring, country $i$'s four branch payoffs are cooperative value $\mathcal C_i$, breakout $\mathcal T_i$ (an advance the rival does not answer), contested race $\mathcal R_i$, and waiting value $\mathcal S_i$ while the rival advances. The outcomes are
\begin{center}
\begin{tabular}{@{}lcc@{}}
\toprule
& the follower pauses & the follower advances\\
\midrule
leader pauses & $(\mathcal C_A,\mathcal C_B)$ & $(\mathcal S_A,\mathcal T_B)$\\
leader advances & $(\mathcal T_A,\mathcal S_B)$ & $(\mathcal R_A,\mathcal R_B)$\\
\bottomrule
\end{tabular}
\end{center}
Maintain the prisoners' dilemma ordering $\mathcal T_i>\mathcal C_i>\mathcal R_i>\mathcal S_i$ for both countries. Each country prefers mutual restraint to mutual advancement, but without monitoring, advancing is strictly dominant. The unique no-agreement outcome is mutual advancement with payoff $\mathcal R_i$.

The four branch payoffs show how survival, strategic control, and costs determine the gains from cooperation. Let $Y_i\ge0$ be country $i$'s value from survival apart from strategic control, and $Z_i>0$ its additional value of control (distinct from the commercial prize in \citealp{gan2026racing}). Let $\chi_i\in(0,1)$ be its share of control in a contested race, with $\chi_A+\chi_B=1$. Active development costs $c_i$ in total, pausing to the end costs $d_i$ with $c_i>d_i\ge0$, and terminal catastrophe probabilities satisfy $0\le p_0<p_1\le p_2<1$, where $p_k$ applies when $k$ countries advance. Catastrophe eliminates $Y_i$ and $Z_i$ after costs are incurred, so
\begin{align}
\mathcal C_i&=(1-p_0)(Y_i+\chi_i Z_i)-d_i,
&\mathcal T_i&=(1-p_1)(Y_i+Z_i)-c_i,\label{eq:game_payoffs}\\
\mathcal R_i&=(1-p_2)(Y_i+\chi_i Z_i)-c_i,
&\mathcal S_i&=(1-p_1)Y_i-d_i.\nonumber
\end{align}
Here each branch payoff multiplies surviving value by one minus that branch's catastrophe probability and subtracts the branch cost.

The gain from cooperation, $\mathcal C_i-\mathcal R_i$, limits what each country will pay for verification.

The leader proposes a common monitoring intensity $m\ge0$ and the follower accepts or refuses. Refusal gives the no-agreement race $\mathcal R_i$. Acceptance requires each country to pay the resource cost $\eta_i m$ up front, where $\eta_i>0$. Countries then simultaneously choose whether to pause or secretly advance.

If only country $i$ advances, a truthful public finding arrives at Poisson rate $m$ per month within the response window $w_i>0$. A timely finding lets the compliant country $j$ choose whether to restart or continue waiting. Restarting restores the contested continuation $(\mathcal R_A,\mathcal R_B)$; waiting leaves $(\mathcal T_i,\mathcal S_j)$, the same outcome as when no timely finding arrives. Mutual compliance yields $\mathcal C_i$, mutual advancement yields $\mathcal R_i$, and $\eta_i m$ is subtracted in every accepted branch.

Catastrophe lotteries are drawn at the end of the block. Interim catastrophe and discounting would make a late finding cost the deviator less, so deterrence would need more monitoring than the threshold derived below. Holding the deviation gain fixed, the dates in this section are then the latest dates for the stated inputs.\footnote{The online supplement, ``General Protocols and Hazard-Weighted Monitoring,'' gives the formula with interim hazard and discounting.}

\subsubsection{Baseline Payoffs}

For the illustrative baseline, set $Y_i=1$, $Z_i=4$, $\chi_i=1/2$, $(p_0,p_1,p_2)=(0.1,0.2,0.4)$, $(c_i,d_i)=(0.8,0.2)$, and $\eta_i=27$ for both countries. No evidence guides $Y_i$ and $Z_i$, so the paper does not calibrate them. They enter the results only through the four payoffs, and the other national values here are illustrative as well.

For example, $\mathcal C=(1-0.1)(1+\tfrac12\times4)-0.2=\agC$, $\mathcal R=(1-0.4)(1+\tfrac12\times4)-0.8=\agR$, and the same formula gives $\mathcal T=\agT$. The gain from cooperation before monitoring is $(p_2-p_0)(Y_i+\chi_iZ_i)+c_i-d_i=0.3\times3+0.6=\rrCminusR{}$ for each country, of which lower catastrophe risk supplies 0.9 and the saved development cost 0.6.

The response example sets $\ell=0$ and separately assumes $w_A=\agUSWindow$ months. The proposed pause lasts \agPauseLength{} months; the remaining interval exceeds catch-up at each displayed review date.

\subsubsection{Credible Response}

\emph{Step 1: the continuation choice.} Take the moment after a timely finding. The upfront monitoring payment is sunk; restarting gives the contested race $\mathcal R_j$, whereas waiting gives $\mathcal S_j$.

\emph{Step 2: benefit versus cost.} By \eqref{eq:game_payoffs}, restarting is credible when the surviving strategic stake exceeds the extra development cost plus the extra catastrophe risk on non-strategic value.

\begin{proposition}[Credible Response]\label{prop:credible_response}
After a truthful timely finding, restarting is strictly optimal for country $j$ if and only if
\begin{equation}
(1-p_2)\chi_j Z_j>
c_j-d_j+(p_2-p_1)Y_j.
\label{eq:credible_response}
\end{equation}
In the maintained prisoners' dilemma region this holds for both countries.
\end{proposition}
The proof is in Appendix~\ref{app:proof_credible_response}.

The left side is the strategic value recovered by restarting, weighted by the chance of survival. The right side is the extra development cost plus the extra expected loss of non-strategic value from a riskier race. Credibility is the $\mathcal R>\mathcal S$ part of the maintained ordering; the proposition shows what that part requires of the primitives. If the inequality fails, so does the ordering, and the threat to restart cannot support the equilibrium below. In the baseline example, the left side is \rrResumeGain{} and the right side is \rrResumeCost, so restarting pays.

\subsubsection{Deterrence and Participation}

Three steps lead to the condition for cooperation.

\emph{Step 1: the gamble.} If the rival complies, a country that secretly advances is caught within its window with probability $1-e^{-mw_i}$. Caught, it gets the contested payoff $\mathcal R_i$; uncaught, the breakout payoff $\mathcal T_i$.

\emph{Step 2: deterrence.} Compliance must be worth at least this gamble. Taking logs, this needs $m\ge\xi_i/w_i$, where $\xi_i$ measures how tempting breakout is relative to cooperation.

\emph{Step 3: participation.} Each country pays $\eta_im$ for monitoring and accepts only if cooperation net of that cost beats the no-agreement race, $\mathcal R_i$. So $m$ can be at most $(\mathcal C_i-\mathcal R_i)/\eta_i$. Both countries must pass both tests.

\begin{proposition}[Costed Monitoring and Voluntary Cooperation]\label{prop:small_agreement}
Under the maintained payoff ordering and response technology, an accepted protocol with mutual compliance can be supported in equilibrium if and only if
\begin{equation}
m_{\min}\equiv\max_{i\in\{A,B\}}\frac{\xi_i}{w_i}
\ \le\ 
m_{\max}\equiv\min_{i\in\{A,B\}}
\frac{\mathcal C_i-\mathcal R_i}{\eta_i},
\qquad
\xi_i\equiv\ln\frac{\mathcal T_i-\mathcal R_i}
{\mathcal C_i-\mathcal R_i}>0.
\label{eq:game_interval}
\end{equation}
Among cooperation-supporting protocols both countries prefer $m^*=m_{\min}$. There is a proposal-and-acceptance equilibrium implementing this intensity, with acceptance and compliance at indifference. Strict deterrence and strict participation are possible if and only if $m_{\min}<m_{\max}$. Mutual advancement remains an equilibrium of the accepted action game, so the result establishes existence, not unique selection of cooperation.
\end{proposition}
The proof is in Appendix~\ref{app:proof_small_agreement}.

The lower bound prices deterrence: a shorter window requires more findings per month to prevent a profitable head start. The upper bound prices consent, because verification cannot consume more than either country's gain from cooperation. So faster detection can help and still be too costly to secure voluntary entry.

The baseline payoffs give
\[
(\mathcal C,\mathcal T,\mathcal R,\mathcal S)=(\agC,\agT,\agR,\agS),
\qquad \xi_i=\agLambda.
\]
Here $\xi_i=\ln(\rrTminusR/\rrCminusR)$ is the breakout temptation, and the participation bound is $m_{\max}=\rrCminusR/27\approx\agMonitoringMaximum$ findings per month.

Deterrence does not require near-certain detection. At date zero the follower's window is $\rrWindowZero$ months, so deterrence needs $m=\rrLambda/\rrWindowZero\approx\rrMinZero$ findings per month. A secret advance is then detected within the window with probability $1-e^{-\xi_i}\approx\rrDetectProb$ percent, and the mean time to a finding is about $\rrMeanDetect$ months. A follower caught advancing falls back to the contested payoff $\mathcal R=\agR$. At this detection probability, the expected payoff of a secret advance equals cooperation's $\mathcal C=\agC$.

Deterrence is an incentive criterion, and at the threshold a breakout still succeeds with probability $e^{-\xi_i}\approx\rrBreakoutProb$ percent. Holding that probability below a tolerated level $\beta$ needs $m\ge\ln(1/\beta)/w_B$, which exceeds the deterrence threshold whenever $\beta$ is below \rrBreakoutProb{} percent.

\subsection{The Agreement Clock}\label{sec:window}

The clock follows from tying the game to the physical response window. For a secret follower advance at pause date $d$, take
\begin{equation}
w_B(d)=\max\{T_{\mathrm{catch}}(d)-\ell,0\},
\label{eq:game_physical_window}
\end{equation}
where $\ell$ is the time needed to restart. This uses catch-up as the boundary beyond which waiting loses the stipulated strategic position. A timely alarm is valuable only if the response actually restores the contested payoff. The window for a secret advance by the leader, $w_A(d)$, must be specified separately; the follower ODE does not provide a symmetric window. When the leader's window binds, it decides feasibility even if the follower's window is long. So the game is a small benchmark tied to one physical window, the follower's. Section~\ref{sec:symmetric} lets the leader's window shrink with compute as well.

The window also carries an assumption about what makes a breakout decisive. It treats being overtaken at the frozen frontier as the loss of the strategic position, and the physical model does not supply that loss by itself. In the calibration, a follower that advances from the start of the pause while the leader waits the full 36 months overtakes the frozen frontier, yet the leader still reaches TED-AI first, at month \mcLeaderFinish, after its scheduled restart. Once the follower is ahead it loses diffusion, and its own research runs at \rrOwnRatio{} of the leader's pace at the same capability and date. So the game describes stakes in which a temporary capability lead is valuable in itself, for example because it can be used before either side reaches TED-AI. If decisiveness meant arriving first, a follower breakout would not be decisive in this calibration.

This breakout interpretation applies when the remaining pause exceeds catch-up time. If the scheduled restart occurs first, the breakout payoff $\mathcal T_i$ is inappropriate and must be replaced by that earlier continuation. The physical condition \eqref{eq:physical_response} handles scheduled expiry; the present game's simple payoff table does not price every possible expiry path.

The corollary follows in three steps.

\emph{Step 1: a floor on the window.} Proposition~\ref{prop:small_agreement} needs $\xi_B/w_B\le m_{\max}$. So the follower's window must be at least $\xi_B/m_{\max}$, the critical window.

\emph{Step 2: a floor on catch-up time.} The window is catch-up time less the implementation delay, by \eqref{eq:game_physical_window}. So catch-up time must be at least $\ell$ plus the critical window.

\emph{Step 3: catch-up time falls.} During a compliant pause both frontiers stay frozen while compute grows. A secret program launched later starts from the same states with more research capacity, so it reaches the frozen leader sooner. Catch-up time falls toward zero, and the floor fails at late enough dates.

\begin{corollary}[A Shrinking Window Can Eliminate Cooperation]\label{cor:game_window}
Hold national payoffs, monitoring costs, implementation delay, and the other country's nonbinding window fixed. At a review date with enough remaining pause for unopposed breakout, mutual compliance is supportable only if
\begin{equation}
T_{\mathrm{catch}}(d)\ge\ell+\frac{\xi_B}{m_{\max}}.
\label{eq:game_window_bound}
\end{equation}
During a compliant pause with fixed frontier states, $a_B>0$, $\gamma_B>0$, and the maintained $\delta\ge0$, $T_{\mathrm{catch}}(d)$ is strictly decreasing and converges to zero. So the set of review dates satisfying this condition is bounded, even though the compliant capability gap has not narrowed.
\end{corollary}
The proof is in Appendix~\ref{app:proof_game_window}.

A country can retain the same visible capability lead yet lose time to act on a secret advance. The required monitoring investment rises as that time disappears and, with national values and monitoring costs fixed, eventually exhausts the gains that finance cooperation. Unlike the shifts in power in \citet{powell2006}, the visible capability gap here does not move, while latent breakout capacity grows with installed compute, so the visible gap is a poor measure of the time to respond.

In the example the two bounds meet, $\xi_i/w=m_{\max}$, at a critical window of $w=\rrLambda/\agMonitoringMaximum\approx\agCriticalWindow$ months. The follower's shrinking window sets the monitoring requirement after date zero (Table~\ref{tab:agreement_game}). At month 18, the least costly deterrent intensity is \agEighteenMonitoring{} and net cooperation pays \agEighteenNetCooperation{}, above the outside payoff $\mathcal R=1$. At month 24, the required intensity is \agTwentyFourMonitoring{} and net cooperation pays only \agTwentyFourNetCooperation{}. Deterrence is technically possible there, but both countries refuse its cost in this example.

\begin{figure}[htbp]
\centering
\includegraphics[width=\textwidth]{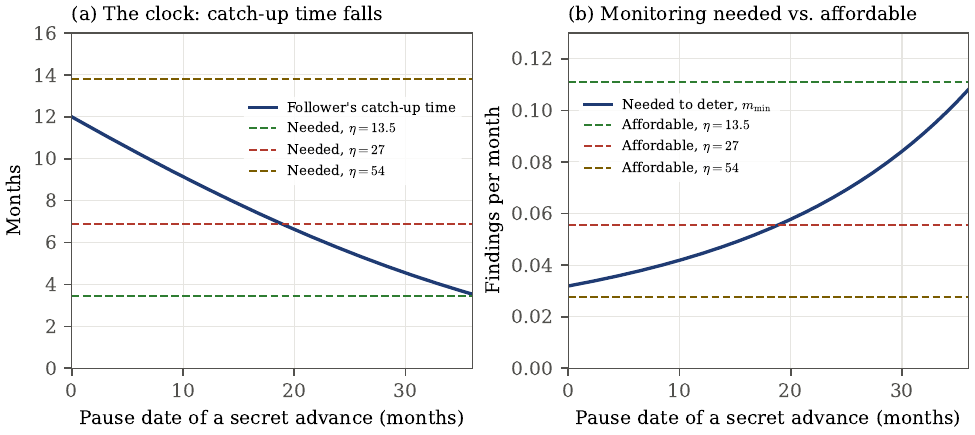}
\caption{The agreement clock. (a) The follower's catch-up time after a secret advance at each pause date, against the catch-up time that the deterrence and participation bounds need at three monitoring costs $\eta$, with no implementation delay. (b) The monitoring intensity needed to deter, $m_{\min}$, against the most both countries will pay, $m_{\max}$. National values are those of the solved example; the leader's window is fixed at 12 months.}
\label{fig:clock}
\end{figure}

The clock comes from the physics; the date comes from chosen inputs. The physical model fixes the catch-up path, but the critical window combines the national values and the monitoring cost, $\xi_i/m_{\max}$. Figure~\ref{fig:clock} and Table~\ref{tab:clock} vary the monitoring cost and the implementation delay. At the baseline cost, cooperation is supportable in review blocks up to month \ackLastBaseZero; one or two months of implementation delay move that date to \ackLastBaseOne{} and \ackLastBaseTwo. Halving the cost keeps cooperation supportable through month \ackGameRangeEnd. After that date the remaining pause is shorter than the catch-up time, so the scheduled restart comes first and the game no longer applies. Doubling it rules cooperation out from the start, because the required monitoring then exceeds what either country will pay even before the window shrinks. 

The online supplement shows how the clock depends on physical and strategic inputs. Table S1 varies the diffusion rate, compute growth, the monitoring cost, the discount rate, and the takeoff length one at a time; compute growth and the monitoring cost move the date most. With compute doubling every 3 months instead of 6, the last supportable review falls from month \ackLastBaseZero{} to \crbGrowthThreeLast; the takeoff length hardly moves it.

So an agreement's duration cannot be read off the capability lead alone, and between rivals that are more alike the clock runs faster (Section~\ref{sec:symmetric}). Monitoring that becomes cheaper, or a gain from cooperation that grows, could also offset the shrinking window; Corollary~\ref{cor:game_window} holds those inputs fixed. Section~\ref{sec:matched_safety} gives one term that grows the gain from cooperation.

\begin{table}[htbp]\centering\small
\caption{When the Agreement Clock Runs Out}\label{tab:clock}
\begin{tabular}{@{}lcccc@{}}\toprule
Monitoring cost & Implementation & Catch-up time & Supportable & Last supportable\\
per unit, $\eta$ & delay $\ell$ (mo) & needed (mo) & at the start? & review date (mo)\\ \midrule
13.5 & 0 & 3.45 & yes & 31.8$^{\dagger}$\\
13.5 & 1 & 4.45 & yes & 30.6\\
13.5 & 2 & 5.45 & yes & 25.4\\
27 & 0 & 6.89 & yes & 18.9\\
27 & 1 & 7.89 & yes & 14.8\\
27 & 2 & 8.89 & yes & 10.9\\
54 & 0 & 13.79 & no & never\\
54 & 1 & 14.79 & no & never\\
54 & 2 & 15.79 & no & never\\
\bottomrule\end{tabular}
\par\smallskip\begin{minipage}{\textwidth}\footnotesize
\emph{Notes:} National values as in the solved example; the leader's window is fixed at 12 months. Cooperation is supportable in the date-$d$ review block when the follower's catch-up time is at least the implementation delay plus $\xi_i/m_{\max}$ and the leader's window also passes (it fails when $\eta>\ackEtaLeaderMax$). $^{\dagger}$Supportable at every review date through month \ackGameRangeEnd, when the remaining 36-month pause falls to the catch-up time. Later, the scheduled restart comes before a breakout could succeed, and the game does not apply. The last date is a conditional boundary for the stated inputs, not an estimated treaty lifetime.
\end{minipage}\end{table}

\begin{table}[htbp]
\centering\small
\caption{Response Windows and Cooperation in the Solved Agreement Example}\label{tab:agreement_game}
\begin{tabular}{@{}rrrrc@{}}
\toprule
Pause date & Follower's window & Minimum $m$ & Net cooperation & Supportable?\\
(months) & (months) & (per month) & payoff & \\
\midrule
0 & 12.00 & 0.0319 & 1.638 & Yes\\
6 & 10.25 & 0.0374 & 1.491 & Yes\\
12 & 8.61 & 0.0445 & 1.299 & Yes\\
18 & 7.10 & 0.0539 & 1.044 & Yes\\
24 & 5.74 & 0.0667 & 0.699 & No\\
\bottomrule
\end{tabular}
\par\smallskip
\begin{minipage}{.94\textwidth}\footnotesize
\emph{Notes:} Illustrative utility primitives give $(\mathcal C,\mathcal T,\mathcal R,\mathcal S)=(2.5,3.2,1.0,0.6)$ and monitoring cost $27m$ per country. Net cooperation is $\mathcal C-27m_{\min}$; the no-agreement payoff is $\mathcal R=1$. The follower's window uses the finest-step physical catch-up calculation with zero implementation delay. The leader's window is separately set to 12 months. A proposed 36-month pause leaves enough time for unopposed breakout at every displayed date. National utilities, monitoring costs, and terminal branch probabilities are chosen game inputs, not estimates or the simulated takeoff probabilities.
\end{minipage}
\end{table}

\FloatBarrier
\subsection{When the Two Countries Are More Alike}\label{sec:symmetric}

The more alike the two countries are, the less time each has to answer a secret advance, so an agreement between near equals needs more monitoring and ends sooner. The main example has a clear leader and follower. The follower has less compute and is six months behind in capability, while the two countries have equal stakes in a contested race. ``More alike'' can mean closer compute or equal capability, and these do not work the same way. Table~\ref{tab:symmetric} takes them one at a time and also relaxes two assumptions of the main example that bear on the same question, equal stakes and the fixed leader window; Appendix~\ref{app:symmetric} gives the details. The binding window is the window of the country whose deterrence needs more monitoring. Cooperation in a review block is supportable while that window is at least the critical window, \symCritical{} months.

Closer compute shortens the window. Panel 1 keeps the six-month lag, lowers the follower's diffusion and raises its own research so that it keeps its initial speed. As diffusion's share of the follower's initial progress falls from \symCopyBaseShare{} to \symCopyHalfShare{} percent, its catch-up time at the start of the pause falls from \symCopyBaseStart{} to \symCopyHalfStart{} months, and the last supportable review moves from month \symCopyBaseLast{} to \symCopyHalfLast. A follower that keeps pace through own research alone catches up in \symCopyNoneStart{} months, less than the critical window, so cooperation is never supportable. Diffusion fades as the gap closes, which is why the calibrated follower's window is twice its lag. Own research does not fade, and it grows with compute.

\begin{table}[htbp]\centering\small\setlength{\tabcolsep}{5pt}
\caption{The Clock When the Rivals Are More Alike}\label{tab:symmetric}
\begin{tabular}{@{}lccccc@{}}\toprule
 & \multicolumn{3}{c}{Binding window (mo)} & Critical & Last supportable\\
\cmidrule(lr){2-4}
 & at the start & month 12 & month 24 & window (mo) & review date (mo)\\ \midrule
\multicolumn{6}{@{}l}{\emph{1. Closer in compute, six-month lag kept: diffusion share; own research relative to the leader's}}\\
81\%; 0.13 (baseline) & 12.0 & 8.61 & 5.74 & 6.89 & 18.9\\
61\%; 0.27 & 9.1 & 5.91 & 3.55 & 6.89 & 8.0\\
41\%; 0.42 & 7.6 & 4.61 & 2.60 & 6.89 & 2.5\\
0\%; 0.71 & 6.0 & 3.26 & 1.71 & 6.89 & never\\
0\%; 1.00 (equal compute) & 4.4 & 2.37 & 1.22 & 6.89 & never\\
\addlinespace
\multicolumn{6}{@{}l}{\emph{2. Equal in capability and compute: a lead of $\Delta$ months is decisive}}\\
$\Delta=0$ (rival passes your frontier) & 0.0 & 0.00 & 0.00 & 6.89 & never\\
$\Delta=6$ & 6.0 & 3.26 & 1.71 & 6.89 & never\\
$\Delta=9$ & 9.0 & 5.08 & 2.72 & 6.89 & 5.7\\
$\Delta=12$ & 12.0 & 7.02 & 3.86 & 6.89 & 12.4\\
$\Delta=18$ (the whole takeoff period) & 18.0 & 11.24 & 6.52 & 6.89 & 22.8\\
\addlinespace
\multicolumn{6}{@{}l}{\emph{3. Unequal stakes: the leader's share of control in a contested race, $\chi_A$}}\\
0.45 & 12.0 & 12.00 & 5.74 & 5.39 & 25.7\\
0.50 (baseline) & 12.0 & 8.61 & 5.74 & 6.89 & 18.9\\
0.55 & 12.0 & 8.61 & 5.74 & 8.94 & 10.8\\
0.60 & 12.0 & 8.61 & 5.74 & 11.15 & 2.9\\
\addlinespace
\multicolumn{6}{@{}l}{\emph{4. The leader's window}}\\
Fixed at 12 months (baseline) & 12.0 & 8.61 & 5.74 & 6.89 & 18.9\\
Shrinks with compute, $\Delta=12$ & 12.0 & 7.02 & 3.86 & 6.89 & 12.4\\
Shrinks with compute, $\Delta=18$ & 12.0 & 8.61 & 5.74 & 6.89 & 18.9\\
\bottomrule\end{tabular}
\par\smallskip\begin{minipage}{\textwidth}\footnotesize
\emph{Notes:} Monitoring cost $\eta=27$, no implementation delay, the 36-month pause and the game's national values, except the control shares in Panel 3. The binding window is the window of the country whose deterrence needs more monitoring; with equal stakes it is the shorter one. The critical window is that of the country that binds at the last supportable date. Panel 1 keeps the six-month lag and the follower's initial speed; its last row gives the follower the leader's research capacity and no diffusion. Panels 2 and 4 measure a window as the time a secret full-speed program needs to move as far as the full-speed path moves in the first $\Delta$ months of the takeoff period, \eqref{eq:tau}. Panels 1, 3 and 4 keep the follower's catch-up time as its window; Panels 1 and 3 keep the leader's 12-month window. The dates are conditional boundaries for the stated inputs, not estimated treaty lifetimes.
\end{minipage}\end{table}

Equal capability removes the window as Section~\ref{sec:window} defines it. With both frontiers paused at the Automated Coder, a secret advance puts the deviator ahead at once, so no finite monitoring deters it. Between equals, what a breakout must achieve is a decisive lead. Suppose a breakout is decisive once a secret full-speed program moves as far as the full-speed path moves in the first $\Delta$ months of the takeoff period. The window at pause date $d$ is then
\begin{equation}
\tau_\Delta(d)=\frac1\gamma\ln\!\left[1+\bigl(e^{\gamma\Delta}-1\bigr)e^{-\gamma d}\right],
\label{eq:tau}
\end{equation}
which equals $\Delta$ at the start of the pause and falls toward zero as compute grows. Both countries face this window, so cooperation in the date-$d$ review block is supportable until
\begin{equation}
d^*_\Delta=\frac1\gamma\ln\frac{e^{\gamma\Delta}-1}{e^{\gamma\xi/m_{\max}}-1}.
\label{eq:symmetric_date}
\end{equation}
The numerator, $e^{\gamma\Delta}-1$, is how much research capacity grows over $\Delta$ months, and the denominator is the same growth over the critical window. So a longer decisive lead or a shorter critical window moves the date later, and cooperation can start only when the decisive lead is at least the critical window.

A decisive lead of six months is never enough. With $\Delta=12$, cooperation is supportable until month \symLeadTwelveLast{}, against month \symCopyBaseLast{} in the main example. By month 24 the window between equals is \symLeadTwelveAtTwentyFour{} months, against \symCopyBaseAtTwentyFour{} for the calibrated follower, because a secret program between equals runs on own research alone.

Stakes and the leader's own window move the date as well. If the leader would win 55 percent of control in a contested race, the follower gains less from cooperating, so a breakout tempts it more, and the date falls to month \symStakesFiftyFiveLast. If the leader's window also shrinks as compute grows and a leader breakout is decisive after 12 months of full-speed progress, the leader's window binds first and brings the date to month \symLeaderTwelveLast. If a leader breakout is decisive only when it completes the takeoff period, the date stays at month \symLeaderEighteenLast{} (Table~\ref{tab:symmetric}, Panel 4).

Monitoring matters more between near equals. The monitoring needed to deter a breakout is inversely proportional to the window (Proposition~\ref{prop:small_agreement}). A follower with the leader's compute, still six months behind, catches up in \symCopyEqualStart{} months at the start of the pause, against \symCopyBaseStart{} in the main example, so deterring it takes about \symMonitorRatio{} times as much monitoring. Cheaper monitoring and a faster response, which lengthen the clock in the main example (Table~\ref{tab:clock}), decide between near equals whether an agreement can start at all. At half the baseline cost the catch-up time needed falls to \ackCritLowZero{} months (Table~\ref{tab:clock}), below the \symCopyEqualStart{} months of a follower with the leader's compute, so an agreement that could not start between them now can. In every row of Table~\ref{tab:symmetric} with a positive window, the window is shorter at month 24 than at the start, so the clock survives each change; what moves is the date.

\FloatBarrier
\subsection{Matched Safety Research as a Verified Term}\label{sec:matched_safety}

An agreement can lengthen its own clock. Beyond stopping capability research, it can commit both countries to the same safety research and verify that commitment along with the pause.

Take the strongest version of this term. Under compliance, the follower does as much safety research as the leader, so its safety knowledge matches the leader's at every date ($S_B=S_A$). The term binds only while both countries comply. With no agreement, or after a breakout, the follower returns to its own program. So only the cooperative branch becomes safer. Safety does not enter the progress law \eqref{eq:china_progress}, so the follower's catch-up path, and with it the physical clock, is unchanged.

The physical model gives the size of the risk cut. With matched safety research, a compliant 36-month mutual pause lowers the global catastrophe probability from \vsPaperThirtySix\% to \vsMatchedThirtySix\% under first-arrival closure, a cut of \vsCutThirtySix\%. The cut is \vsCutTwelve\% for a 12-month pause and \vsCutTwentyFour\% for a 24-month pause.

The game turns this cut into time in three steps.

\emph{Step 1: cooperation pays more.} A lower cooperative-branch risk $p_0$ raises the cooperative payoff $\mathcal C$ in \eqref{eq:game_payoffs} and leaves $\mathcal T$, $\mathcal R$, and $\mathcal S$ unchanged.

\emph{Step 2: breakout tempts less, and monitoring is worth more.} A higher $\mathcal C$ lowers the breakout temptation $\xi$ and raises the most both countries will pay for monitoring, $m_{\max}=(\mathcal C-\mathcal R)/\eta$.

\emph{Step 3: a shorter window suffices.} Both changes lower the critical window $\xi/m_{\max}$, so cooperation stays supportable at later review dates, when catch-up time is shorter.

In the game, $p_0=0.1$ is a chosen input, not a simulated probability. As an illustration, apply the simulated \vsCutThirtySix\% cut to it, which gives $p_0=\vsPzeroScaled$. Cooperation then pays $\mathcal C=\vsCScaled$ instead of \agC, the critical window falls from \agCriticalWindow{} to \vsCritScaled{} months, and cooperation stays supportable through month \vsLastScaled{} instead of month \ackLastBaseZero. Each percentage point cut in $p_0$ adds \vsPerPointLow--\vsPerPointHigh{} months; at $p_0=0.05$, cooperation stays supportable through month \vsLastPzeroFive.

Verifying a safety program has a cost too. The gain survives as long as the monitoring cost per unit stays below \vsEtaBreakEven, which is \vsEtaBreakEvenPct\% above the baseline of 27. If the inspections that verify a pause also cover the safety program, the added cost can be small. So an agreement that makes cooperation safer also makes cooperation last longer.

\subsection{The Scenario's Own Risks}\label{sec:own_risks}

The game's branch risks describe the stakes as governments perceive them, and they are milder than the scenario's own. Along the scenario's paths, cooperation through the 36-month pause carries \mcCoop{} percent and a race from the start of the pause \mcRaceZero{} percent. At that risk the leader would rather keep waiting than answer a breakout, so its threat to restart is not credible (Proposition~\ref{prop:credible_response}) and no monitoring can deter. Safety research during the pause makes a race survivable, and the threat becomes credible after \mcOpening{} months. At the baseline monitoring cost, cooperation is supportable from month \mcOpening{} to month \mcCloseBase. So at the baseline hazard level an agreement has an opening date as well as a closing date. The online supplement reports the dates at other monitoring costs (Table S2). At a full-speed probability of 25 percent a race carries \mcRaceZeroLo{} percent, so the restart is credible from the first date, but cooperation is worth less. At the baseline monitoring cost the agreement is supportable only through month \mcCloseBaseLo{}, and at half the cost through month \mcCloseHalfLo{} (Table S3).

\section{Widening the Lead and Ranking Instruments}\label{sec:results}

Sections~\ref{sec:china} and~\ref{sec:agreement} give the problem and the agreement. This section asks which levers move them. Cutting the follower's own research or diffusion lengthens catch-up time and so widens the response window. Domestic pacing rules can instead raise global risk, because they let the follower pull alongside.

\subsection{Widening the Lead: Own Research and Diffusion}\label{sec:lead_measures}

Policy can lengthen catch-up time by cutting the follower's own research or diffusion. The constant-speed result compares changes in $a$ and $\delta$; the dynamic scenarios below separately vary own productivity, research-capacity growth, and diffusion. Catch-up time $T^*=\ln(1+u)/\delta$ responds to the two cuts in different ways.

\emph{Step 1: own research.} Catch-up time depends on $a$ only through $u=\delta G_0/a$. For a small change, the gain per unit reduction in $\ln a$ is $G_0/(a+\delta G_0)$ months: the raw lead.

\emph{Step 2: diffusion.} A small reduction in $\ln\delta$ has two effects. Per unit reduction, slower diffusion adds $T^*$ months, while the lower diffusion-to-research ratio $u$ takes back the raw lead. The net gain is $T^*$ less the raw lead.

\begin{proposition}[Comparative Statics of Catch-Up Time]\label{prop:buy}
The catch-up time $T^*$ satisfies the sensitivities:
\begin{equation}
\frac{\partial T^*}{\partial \ln a} = -\,\text{raw lead}, \qquad \frac{\partial T^*}{\partial \ln \delta} = -\,\bigl(T^* - \text{raw lead}\bigr).
\label{eq:sensitivities}
\end{equation}
\end{proposition}

The proof is in Appendix~\ref{app:proof_buy}.

Take a catch-up time of twice the raw lead ($T^* = 2\,\text{raw}$). The two semielasticities, measured in months, then have the same size. With the 12-month catch-up time and a 6-month raw lead, a 10\% cut in own research and a 10\% cut in diffusion each add about 0.6 months.

\begin{table}[htbp]
\centering
\caption{Widening the Geopolitical Lead}
\label{tab:lead}
\small
\setlength{\tabcolsep}{5pt}
\begin{tabular}{@{}>{\raggedright\arraybackslash}p{5.2cm}ccccc@{}}
\toprule
& Catch-up & Longest & \multicolumn{3}{c}{Global catastrophe probability (\%)}\\
\cmidrule(l){4-6}
Policy intervention & time (mo) & leader pause (mo) & Baseline & Own longest & Gain from\\
& $T^*$ & $L^*$ & pause & pause & longer pause\\
\midrule
Baseline (no intervention) & \cnCatchUp & \cnMaxPause & \cnMaxPauseLeader & \cnMaxPauseLeader & --\\
Follower research-capacity growth halved ($\gamma_B / 2$) & \cnEcGrowthCatchUp & \cnEcGrowthPause & \cnEcGrowthSame & \cnEcGrowthLeader & \cnEcGrowthLonger\\
Follower own speed cut by 25\% & \cnEcLevelCatchUp & \cnEcLevelPause & \cnEcLevelSame & \cnEcLevelLeader & \cnEcLevelLonger\\
Diffusion rate halved ($\delta / 2$) & \cnEcDistillCatchUp & \cnEcDistillPause & \cnEcDistillSame & \cnEcDistillLeader & \cnEcDistillLonger\\
\bottomrule
\end{tabular}
\par\smallskip
\begin{minipage}{\textwidth}
\footnotesize
\emph{Notes:} Computed under first-arrival closure. Changes in research-capacity growth and own speed stylize hardware restrictions; changes in diffusion stylize algorithmic security. The common comparison holds the pause at 12 months; the own-longest column extends it to each scenario's lead-preserving limit. Intervention costs and feasibility are not modeled.
\end{minipage}
\end{table}

Table~\ref{tab:lead} reports scenario changes in the full dynamic model. Halving the follower's research-capacity growth lengthens catch-up time by \cnEcGrowthGain{} months, lowering risk by \cnEcGrowthTotal{} points. Halving diffusion lengthens catch-up time by \cnEcDistillGain{} months and lowers the global catastrophe probability by \cnEcDistillTotal{} points (\cnEcDistillDirect{} directly and \cnEcDistillLonger{} through the longer pause). These are different primitive interventions, because a growth reduction leaves initial own productivity unchanged. Both channels can widen the room for restraint.

\subsection{Instrument Rankings when the Follower Races}\label{sec:rankings}

We evaluate the pacing specifications of Table~\ref{tab:policies} when the follower is racing in parallel. A compute floor fixes the research and safety shares for the whole takeoff period and gives the rest to customers. A safety floor raises the safety share and takes it from customers. A risk ceiling slows development just enough to hold the hazard at the ceiling and moves the freed compute into safety. An R\&D-model lag makes research use the model from a fixed number of months earlier. A pause with all compute to customers sets research and safety to zero for its duration; a pause with safety moves the research share into safety instead. 

The table reports both closure rules. To attribute global probability to national origins, let $\Lambda_j(t)=\int_0^t\lambda_j(v)\,dv$ and define cause-specific catastrophe incidence
\begin{equation}
I_j(T)=\int_0^T e^{-\Lambda_A(t)-\Lambda_B(t)}\lambda_j(t)\,dt,
\qquad I_A(T)+I_B(T)=1-e^{-\Lambda_A(T)-\Lambda_B(T)}.
\label{eq:cause_incidence}
\end{equation}
Here $I_j(T)$ weights frontier $j$'s instantaneous hazard $\lambda_j(t)$ by the probability $e^{-\Lambda_A(t)-\Lambda_B(t)}$ that neither frontier has caused a catastrophe before date $t$. Because the two frontiers partition the first catastrophe event, $I_A(T)$ and $I_B(T)$ add up to the global catastrophe probability, whereas standalone national probabilities do not.

The common-share scenarios connect domestic coordination \citep{gan2026racing} to the foreign constraint.\label{sec:synthesis} Without a foreign rival, common-share coordination chooses $\dvXCoop\%$ research and gives a $\dvRiskCoop\%$ domestic catastrophe probability, but that share allows foreign overtaking after about \rrOvertakeCoop{} months. The monthly discount rate is $r=\dvRMonthPct\%$. Retaining baseline safety $s_0$, consider choosing a constant domestic share to minimize global cumulative hazard plus discounting subject to keeping the lead:
\begin{equation}
\min_{x\in[\underline x,1-s_0]}
\left\{H_{\mathrm{global}}(x)+rT_A(x)\right\}
\quad\text{subject to}\quad
F(t;x)>F_B(t;x)\quad\text{until first arrival}.
\label{eq:global_benchmark}
\end{equation}
Here $T_A(x)$ is the leader's arrival date and $\underline x=10^{-4}$ is the computational lower bound. Minimizing \eqref{eq:global_benchmark} is equivalent to maximizing discounted global survival among constant shares, subject to keeping the lead until first arrival (a weaker requirement than the continuing-horizon floor \eqref{eq:boundary_speed}). The smallest common share that keeps the lead through leader completion is $\rrLeadBoundary\%$, with a $\cnCoopLeader\%$ global catastrophe probability. Minimizing \eqref{eq:global_benchmark} selects $\cnCoopGX\%$ research and lowers global catastrophe probability to $\cnCoopGLeader\%$ ($\cnCoopGIncUS\%$ from the leader frontier and $\cnCoopGIncC\%$ from the follower's). The compute floor with 5\% research and 25\% safety achieves $\cnBestFloorLeader\%$. It keeps a higher safety share and uses more research than this optimum, so its advantage comes from safety.

\begin{table}[htbp]
\centering
\caption{Catastrophe Probability with the Follower Racing}
\label{tab:policies}
\small
\setlength{\tabcolsep}{3pt}
\begin{tabular}{@{}>{\raggedright\arraybackslash}p{4.7cm}cccccc@{}}
\toprule
& & & \multicolumn{3}{c}{First-arrival closure (baseline)} & Independent\\
\cmidrule(lr){4-6}
Policy specification & Keeps & Min.\ catch-up & Global & From leader & From follower & closure:\\
& lead & time & (\%) & frontier & frontier & global (\%)\\
\midrule
Full-speed race & yes & \cnBaseCap & \cnBaseLeader & \cnBaseIncUS & \cnBaseIncC & \cnBaseBoth\\
Longest leader pause w/ safety (\cnMaxPause{} mo) & yes & \cnMaxPauseCap & \cnMaxPauseLeader & \cnMaxPauseIncUS & \cnMaxPauseIncC & \cnMaxPauseBoth\\
Compute floors, best share (\cnBestFloor\%) & yes & \cnBestFloorCap & \cnBestFloorLeader & \cnBestFloorIncUS & \cnBestFloorIncC & \cnBestFloorBoth\\
Compute floors, lead boundary (\rrLeadBoundary\%) & yes & \cnMinFloorCap & \cnMinFloorLeader & \cnMinFloorIncUS & \cnMinFloorIncC & \cnMinFloorBoth\\
Pause, all to customers (12 mo) & no & \cnSPauseBlogCap & \cnSPauseBlogLeader & \cnSPauseBlogIncUS & \cnSPauseBlogIncC & \cnSPauseBlogBoth\\
Monthly risk ceiling & yes & \cnSCeilingCap & \cnSCeilingLeader & \cnSCeilingIncUS & \cnSCeilingIncC & \cnSCeilingBoth\\
R\&D-model lag & yes & \cnSLagCap & \cnSLagLeader & \cnSLagIncUS & \cnSLagIncC & \cnSLagBoth\\
Safety floor (25\% from customers) & yes & \cnSSafetyFloorCap & \cnSSafetyFloorLeader & \cnSSafetyFloorIncUS & \cnSSafetyFloorIncC & \cnSSafetyFloorBoth\\
\midrule
Common research share, lead boundary (\rrLeadBoundary\%) & yes & \cnCoopCap & \cnCoopLeader & \cnCoopIncUS & \cnCoopIncC & \cnCoopBoth\\
Common research share, global survival (\cnCoopGX\%) & yes & \cnCoopGCap & \cnCoopGLeader & \cnCoopGIncUS & \cnCoopGIncC & \cnCoopGBoth\\
\bottomrule
\end{tabular}
\par\smallskip
\begin{minipage}{\textwidth}
\footnotesize
\emph{Notes:} Scenario probabilities under first-arrival and independent closure. Minimum catch-up time is the lowest catch-up time along the leader path (12.0 months initially). The ``from leader'' and ``from follower'' columns are additive cause-specific incidences \eqref{eq:cause_incidence}, not standalone national probabilities or marginal effects. The common-share rows retain baseline safety and solve the lead-constrained problem \eqref{eq:global_benchmark}; the compute-floor rows use 25\% safety. Lead-boundary rows use the smallest share that keeps the lead; their minimum catch-up time is near zero, so they keep the lead with no time left to respond.
\end{minipage}
\end{table}

On the leader's own frontier, with the follower ignored, the selected risk ceiling and R\&D-model lag lower the catastrophe probability from \cnBaseUS\% to \cnSCeilingUS\% and \cnSLagUS\%. With the follower racing, they raise the global catastrophe probability instead, from \cnBaseLeader\% in the race to \cnSCeilingLeader\% and \cnSLagLeader\% under first-arrival closure. So a domestic ranking need not survive foreign exposure.

Ceilings and lags fail here because they slow development late in the takeoff period, when capability and the monthly hazard are high. The follower's unpaced frontier can then pull alongside the leader at maximum capability. The follower's contribution to the global catastrophe probability (column 6) rises from \cnBaseIncC\% in the race to \cnSCeilingIncC\% under the ceiling and \cnSLagIncC\% under the lag.

\begin{figure}[htbp]
\centering
\includegraphics[width=\textwidth]{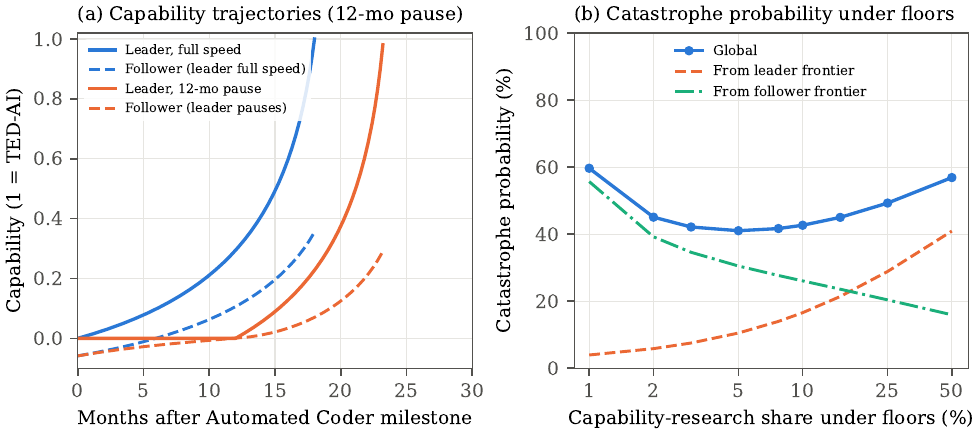}
\caption{(a) Capability trajectories under full speed and a 12-month leader pause with safety research. Diffusion slows as the gap closes; the leader's scheduled restart uses accumulated compute. (b) Global catastrophe probability under compute floors against capability research share, with additive leader and follower cause-specific incidence.}
\label{fig:us_china}
\end{figure}

Compute floors (Figure~\ref{fig:us_china}b) distribute delay across the trajectory. The selected \cnBestFloor\% research share gives a \cnBestFloorLeader\% global catastrophe probability, with \cnBestFloorIncUS\% incidence originating at the leader frontier and \cnBestFloorIncC\% at the follower's. A lower lead-preserving share ($\rrLeadBoundary\%$) extends the takeoff period to \cnMinFloorTus{} months and leaves \cnMinFloorIncC\% incidence from the follower. More domestic delay can raise foreign exposure even before overtaking occurs.

\subsection{Robustness to Closure Rules and Safety Knowledge Spillovers}\label{sec:closure}

Closure matters because the leader's arrival at $F=1$ may not end the follower's danger. Table~\ref{tab:closure} examines the independent closure rule, where the follower's takeoff period runs to completion.

\begin{table}[htbp]
\centering
\caption{Global Catastrophe Probability When the Follower's Takeoff Period Runs to Completion (\%)}
\label{tab:closure}
\small
\begin{tabular}{@{}lccc@{}}
\toprule
& Baseline & Equal follower & Follower adopts half\\
Policy specification & ($\sigma_B=\sigma/2$) & safety ($\sigma_B=\sigma$) & of leader safety progress\\
\midrule
Full-speed race & \cnBaseBoth & \cnPhiOneBase & \cnSpillBase\\
Longest leader pause with safety & \cnMaxPauseBoth & \cnPhiOnePauseBoth & \cnSpillPauseBoth\\
Compute floors (\cnClosureFloorX\% research share) & \cnSFloorsBoth & \cnPhiOneFloorsBoth & \cnSpillFloorsBoth\\
\bottomrule
\end{tabular}
\par\smallskip
\begin{minipage}{\textwidth}
\footnotesize
\emph{Notes:} Cumulative probability when both national takeoff periods run to completion. The floor is the equal-delay setting (\cnClosureFloorX\% research, 25\% safety), distinct from the 5\% floor in Table~\ref{tab:policies}. Follower safety productivity is half the leader baseline in the first column and equal in the second. The third adds adoption of half of leader safety progress to baseline follower learning.
\end{minipage}
\end{table}

With independent closure and baseline follower safety, the global catastrophe probability remains high: the equal-delay floor gives \cnSFloorsBoth\%, against \cnBaseBoth\% without pacing. The 5\% floor in Table~\ref{tab:policies} gives \cnBestFloorBoth\%. These reductions are meaningful but much smaller than under first-arrival closure. Equal own follower safety lowers the equal-delay floor outcome to \cnPhiOneFloorsBoth\%, while adoption of leader safety knowledge lowers it to \cnSpillFloorsBoth\%.

So international safety diffusion is an important channel, but not indispensable in every specification. An agreement can also write it into its terms (Section~\ref{sec:matched_safety}). Slower domestic progress also gives the follower time to accumulate its own safety knowledge. Closure is especially consequential for agreements, as the compliant 24-month mutual pause of Section~\ref{sec:gains} shows.

\FloatBarrier
\section{Scope and Limitations}\label{sec:scope}

The analysis is limited in four respects. First, the foreign-response simulations treat the follower's research and safety program as specified paths. The finite agreement game solves proposal, participation, compliance, and restarting for stipulated national values and costs. Its terminal risk lotteries, committed programs, public truthful findings, and binary restoration of a contested position simplify the continuation. The physical catch-up calculation supplies one response window; it does not derive the strategic allocation or the other country's window. The game example compares review states and holds branch values fixed. It does not solve repeated renewals, renegotiation, a fully strategic investment game, or a joint equilibrium of domestic labs and governments. National utilities and monitoring costs are illustrative inputs, and simulated pause risks assume cooperation. The matched-safety exercise of Section~\ref{sec:matched_safety} carries a simulated risk cut into a chosen input, $p_0$; it shows the size of the effect, not an estimate. 

Within this first respect, the contested share of control is also an input. Contest models derive winning probabilities or shares of control from competing efforts \citep{skaperdas1996,garfinkel2007}. In lottery contests, risk attitudes can affect effort and the odds of control; the effect depends on liability \citep{skaperdas1995}. The branch payoffs here are expected values, so governments are risk neutral over the terminal lotteries. Interim catastrophe and discounting enter the incentive accounting of the online supplement. False findings, private monitoring, and uncertain capability states require richer dynamic structures.

Second, compute is an exogenous growing stock. Pacing can change investment incentives, while endogenous investment can change catch-up time during a treaty. Capability publication also affects international diffusion.

Third, the baseline safety law has share curvature but no scaling with total compute. Different safety productivity, transfer of knowledge, risk shapes, or post-arrival exposure can change the numerical trade-offs. In particular, first-arrival closure assumes the winning frontier terminates the other's risk; independent closure often gives much higher probabilities.

Fourth, the calibration uses uncertain illustrative targets and model-implied diffusion parameters, not identified causal estimates. The numerical checks establish consistency for the specified cases; they do not prove universal policy rankings. Empirical measurements are needed to assess the quantitative scenario.

These omissions cut both ways. Repeated renewal could make agreement easier, because a longer shadow of the future supports cooperation \citep{axelrodkeohane1985}. Better verification technology would also help. False findings, private monitoring, and endogenous investment could make it harder.

\section{Conclusion}\label{sec:conclusion}

Foreign progress shapes domestic restraint. Diffusion slows as a follower approaches a paused leader, so catch-up time exceeds the raw lead, but that room for restraint is finite and state dependent.

Mutual restraint lets both countries build safety knowledge while preserving their relative capability positions. In the game, lower catastrophe risk and development costs give each a gain over mutual racing. That gain can finance verification, but the agreement must also deter a secret advance. Agreements to pause frontier AI are hard, but not out of reach.

Agreements have a clock. Compute keeps growing during a compliant pause, so the time a country has to act on a secret advance shrinks even while the capability gap stays fixed. Deterrence then needs more monitoring. Between rivals that are more alike the window is shorter from the start, so monitoring matters even more. Once that monitoring costs more than one of the countries gains from cooperating, voluntary cooperation is no longer supportable in that review block. This holds national values, monitoring costs, and the implementation delay fixed. When that happens depends on those inputs, not on the capability lead alone.

The clock can be slowed. Cheaper monitoring, a faster response, and matched safety research written into the verified terms all keep cooperation supportable for longer. A longer agreement can also use a detection standard that tightens over time, an earlier scheduled restart, or other safeguards.

Three policy channels do distinct work. First, domestic capability and safety standards can restrain commercial racing and raise safety effort, but a constant research share must stay at about \rrLeadBoundary\% or more to keep the foreign lead. Second, hardware controls and algorithmic security lengthen catch-up time (Table~\ref{tab:lead}), widening both the unilateral pause limit and the response window. Third, verified reciprocal restraint works only if restarting is credible \eqref{eq:credible_response}, the monitoring intensity lies inside the deterrence-and-participation interval \eqref{eq:game_interval}, and catch-up time exceeds the window bound \eqref{eq:game_window_bound}. Each condition names a term that negotiators can write and check.

\appendix
\section{Calibration and Numerical Implementation}\label{app:calibration}

Table~\ref{tab:calibration} separates imposed targets, assumed inputs, fitted parameters, and model implications. This appendix makes the baseline hazard construction and pause target explicit. These calibrations define an illustrative scenario, rather than identify a causal relationship from observed AI outcomes.

Normalize $K_0=1$. The calibration sets $\phi(F)=\theta e^{\rho F}$ with $\rho=\ln10$, and $y(s)=\psi+(1-\psi)(s/s_0)^\alpha$ with $\psi=\alpha=1/2$. At the baseline research share $x_0$, the initial research coefficient and the calendar date at capability $F$ are
\begin{align}
b_0&=\frac{\gamma(1-e^{-\rho})}{\rho(e^{\gamma T_0}-1)},
&\theta&=\frac{b_0}{x_0^\varepsilon},\\
t_0(F)&=\frac1\gamma\ln\left[1+\frac{e^{\gamma T_0}-1}{1-e^{-\rho}}
(1-e^{-\rho F})\right].
\end{align}
Here $b_0$ is the research speed at the start of the full-speed path, and $\theta$ converts it into productivity at the Automated Coder for $K_0=1$. The date $t_0(F)$ is when the full-speed path reaches capability $F$. The full-speed velocity at that capability is $f_0(F)=b_0e^{\gamma t_0(F)+\rho F}$, and baseline safety knowledge is $S_0(F)=\sigma t_0(F)$, starting from zero at $F=0$.

One target fixes the shape of the hazard. Let $q$ be the fraction of full-speed cumulative hazard occurring in the second half of the \emph{capability interval}, $F\in[1/2,1]$. It is not the fraction in the last half of calendar time. With hazard density $a_h e^{c_hF}$ per unit of capability,
\begin{equation}
c_h=2\ln\frac{q}{1-q},\qquad
a_h=\frac{-\ln(1-P_{\rm full})}{\int_0^1 e^{c_hF}\,dF}.
\end{equation}
The shape $c_h$ sets how fast the hazard density rises with capability; with $q=0.75$ it is $2\ln3\approx2.20$. The scale $a_h$ makes full-speed cumulative hazard equal $-\ln(1-P_{\rm full})=\ln4$. For a state-hazard share $\omega$, define the zero-knowledge primitives
\begin{equation}
h(F)=\omega a_h e^{c_hF} f_0(F)e^{S_0(F)},\qquad
k(F)=(1-\omega)a_h e^{c_hF}e^{S_0(F)}.
\end{equation}
Along the reference path, multiplying by $e^{-S_0}$ and using $dt=dF/f_0$ gives $\lambda_0dt=a_h e^{c_hF}dF$. State and step terms contribute fractions $\omega$ and $1-\omega$ of full-speed hazard. Holding these primitives fixed on alternative paths allows accumulated knowledge and the pace of progress to change risk. Follower catastrophe hazard is inactive for $F_B<0$, while safety knowledge continues accumulating; the shared primitives apply once the follower enters $F_B\ge0$.

The imposed pause-risk target suspends capability progress for $L_{\rm cal}$ months at $F=0$, keeps baseline safety research, and stops the compute clock. It then restarts the same reference capability path with the accumulated knowledge. With a constant safety yield, $\sigma$ solves
\begin{equation}
-\ln(1-P_{\rm reg})
=\frac{h(0)}{\sigma}(1-e^{-\sigma L_{\rm cal}})
+e^{-\sigma L_{\rm cal}}[-\ln(1-P_{\rm full})].
\end{equation}
The first term is state exposure during the pause; the second is subsequent hazard reduced by the additional safety stock. The 40\% target is the AI Futures Project's judgment for a safety-case regime \citep{aifutures2026}; assigning it to this specific stopped-clock experiment is a modeling assumption. This experiment is distinct from the accelerating safety-funded pauses and bilateral pauses in the policy tables, which keep installed compute growing and can use different safety shares.

The constant-share global calculation uses the closed-form leader path and adaptive follower and hazard integration. It separates phases at the follower's $F_B=0$ entry and terminal arrivals and detects lead crossing as an event. It searches the finite-horizon feasible share interval for the discounted-survival objective \eqref{eq:global_benchmark}; the 5\% research, 25\% safety floor is a separate policy scenario. Other instruments use time-domain integration with step-size checks.

With a 36-month full-speed takeoff period, the agreement clock barely moves, but pauses remove less risk (Table~\ref{tab:takeoff36}). Thirty-six months is a slow case. In the AI Futures Model's published forecasts for this stretch, the medians run from 9 to 18 months and the 90th percentiles from 3.6 to 15.5 years \citep{lifland2026}. The check keeps the hazard targets, the follower's lag and catch-up targets, the game's payoffs and monitoring cost, and the 36-month pause. It refits the safety yield and both frontiers' progress parameters.

Catch-up time still falls from \tkCatchStartThirtySix{} to \tkCatchTwentyFourThirtySix{} months over 24 months of compliance, so the last supportable review moves only from month \tkLastReviewEighteen{} to \tkLastReviewThirtySix. Doubling the proposed pause to \tkScaledPause{} months extends the game's range to month \tkGameRangeEndScaledThirtySix{} but leaves that date unchanged.

A compliant 36-month mutual pause leaves a global catastrophe probability of \tkMutualThirtySixThirtySix{} percent rather than \tkMutualThirtySixEighteen, and \tkMutualFollowerThirtySix{} points of it come from the follower's frontier, whose safety research is slower. So matched safety research matters more: it brings that probability to \tkMatchedPauseThirtySix{} percent and keeps cooperation supportable to the end of the game's range, month \tkLastMatchedThirtySix{} (month \tkLastMatchedScaledThirtySix{} with the doubled pause). A unilateral leader pause now does best starting at month \tkMaxPauseStartThirtySix{} and leaves \tkMaxPauseLeaderThirtySix{} percent rather than \tkMaxPauseLeaderEighteen.

\begin{table}[htbp]
\centering\small
\caption{Baseline and a 36-Month Takeoff Period}\label{tab:takeoff36}
\begin{tabular}{@{}lrr@{}}
\toprule
& \multicolumn{2}{c}{Full-speed takeoff period $T_0$}\\
\cmidrule(l){2-3}
& 18 months (baseline) & 36 months\\
\midrule
\multicolumn{3}{@{}l}{\emph{Follower calibration}}\\
\quad Diffusion rate $\delta$ (\% of the gap a month) & \tkDeltaEighteen & \tkDeltaThirtySix\\
\quad Initial diffusion share (\%) & \tkCopyShareEighteen & \tkCopyShareThirtySix\\
\addlinespace
\multicolumn{3}{@{}l}{\emph{Agreement clock, $\eta=27$}}\\
\quad Catch-up time at the start of a compliant pause (months) & \tkCatchStartEighteen & \tkCatchStartThirtySix\\
\quad Catch-up time after 24 months of compliance (months) & \tkCatchTwentyFourEighteen & \tkCatchTwentyFourThirtySix\\
\quad Last supportable review date (month) & \tkLastReviewEighteen & \tkLastReviewThirtySix\\
\quad With matched safety research (month) & \tkLastMatchedEighteen & \tkLastMatchedThirtySix$^{\dagger}$\\
\quad End of the game's range (month) & \tkGameRangeEndEighteen & \tkGameRangeEndThirtySix\\
\addlinespace
\multicolumn{3}{@{}l}{\emph{Proposed pause scaled to $2T_0$ (36 and \tkScaledPause{} months)}}\\
\quad With matched safety research (month) & \tkLastMatchedScaledEighteen & \tkLastMatchedScaledThirtySix\\
\quad End of the game's range (month) & \tkGameRangeEndScaledEighteen & \tkGameRangeEndScaledThirtySix\\
\addlinespace
\multicolumn{3}{@{}l}{\emph{Global catastrophe probability (\%)}}\\
\quad Racing & \tkGlobalRaceEighteen & \tkGlobalRaceThirtySix\\
\quad Compliant mutual pause, 24 months & \tkMutualTwentyFourEighteen & \tkMutualTwentyFourThirtySix\\
\quad Compliant mutual pause, 36 months & \tkMutualThirtySixEighteen & \tkMutualThirtySixThirtySix\\
\addlinespace
\multicolumn{3}{@{}l}{\emph{Unilateral leader pause with safety, at its best start}}\\
\quad Best start (month) & \tkMaxPauseStartEighteen & \tkMaxPauseStartThirtySix\\
\quad Longest pause before the follower overtakes (months) & \tkMaxPauseEighteen & \tkMaxPauseThirtySix\\
\quad Global catastrophe probability (\%) & \tkMaxPauseLeaderEighteen & \tkMaxPauseLeaderThirtySix\\
\bottomrule
\end{tabular}
\par\smallskip
\begin{minipage}{\textwidth}\footnotesize
\emph{Notes:} Each column refits the model with the full-speed takeoff period $T_0$ set to 18 or 36 months. Held fixed: the two hazard targets, the compute and research technology, the follower's six-month lag and 12-month catch-up targets, its relative safety yield, the game's payoffs, the monitoring cost $\eta=27$, the 12-month leader window, zero implementation delay, and, in the clock block, the proposed 36-month pause. Refitted: $\sigma$, $b_0$, the follower's own-research speed $a_B$, and $\delta$. In the 36-month case review dates run to month \tkReviewGridEnd, so the date grid cuts off neither date. $^{\dagger}$Supportable at every review date until the game's range ends, when the remaining pause equals the catch-up time. The scaled block lengthens the proposed pause in proportion to the takeoff period, so the game's range covers the longer takeoff period; matched safety then uses the risk cut of that longer mutual pause. The pause search keeps its method but widens its start grid to months 0--\tkPauseStartsEnd{} and its length bracket to \tkPauseCap{} months, in proportion to $T_0$; it picks the start with the lowest global catastrophe probability. Probabilities use first-arrival closure.
\end{minipage}
\end{table}

The replication package records parameter provenance, pause horizons, and date-dependent catch-up times. The game check covers the payoff ordering, optimal responses, equilibria, participation boundaries, and asymmetric cases, and it generates Table~\ref{tab:agreement_game}. National utilities, branch probabilities, and monitoring technology remain additional uncalibrated primitives.

\FloatBarrier
\newpage
\section{Rivals That Are More Alike}\label{app:symmetric}

This appendix gives the details behind Table~\ref{tab:symmetric} and Section~\ref{sec:symmetric}. Every row keeps the monitoring cost at 27, no implementation delay, the 36-month pause and the game's national values, except where the row says otherwise.

Panel 1 brings the follower closer in compute; Section~\ref{sec:symmetric} gives its dates. The panel keeps the six-month capability lag and the condition that the follower keeps pace when the pause begins, and lowers the diffusion rate, so the follower's own research rises to make up the difference. Between the baseline and the half-diffusion row, at a diffusion share of \symCopyThreeQShare{} percent, the follower catches up in \symCopyThreeQStart{} months at the start and the last supportable review is month \symCopyThreeQLast. A follower with no diffusion needs \symCopyNoneOwn{} of the leader's research capacity to keep pace, which is the leader's own capacity six months earlier, or half its compute when $\varepsilon=1/2$. The calibrated follower is far from these cases. In compute terms, its own research equals what \symBaseComputePct{} percent of the leader's compute would give. Without diffusion, the window also falls faster: by month 24 it is \symFallNone{} percent of its value at the start, against \symFallBase{} percent for the calibrated follower, because own research grows with compute and diffusion does not.

Without diffusion, compute and capability enter the window through one ratio. Write the progress law of Section~\ref{sec:model} as $dF/\phi(F)=(xK_0)^\varepsilon e^{\gamma t}\,dt$. Along the leader's full-speed path, the integral of $1/\phi$ over a stretch of capability is then proportional to the growth of $e^{\gamma t}$ over the months the stretch took, whatever the shape of $\phi$. Let the follower be $L$ months of full-speed progress behind, and let its research capacity $\kappa$ be its $(xK)^\varepsilon$ relative to the leader's, a pure ratio. If it advances secretly at pause date $d$ against the frozen leader, it catches up in
\begin{equation}
T(d)=\frac1\gamma\ln\!\left[1+\frac{(1-e^{-\gamma L})\,e^{-\gamma d}}{\kappa}\right].
\label{eq:catch_no_copy}
\end{equation}
Compute, through $\kappa$, and capability, through $L$, enter only as the ratio $(1-e^{-\gamma L})/\kappa$, so one can stand in for the other. Cooperation is supportable at the start of the pause only if $T(0)$ is at least the critical window. That takes a lag of at least \symLagNeededEqual{} months against a follower with the leader's compute, \symLagNeededHalf{} months against one with half of it ($\kappa=\sqrt{1/2}$), and \symLagNeededQuarter{} months against one with a quarter ($\kappa=1/2$). With diffusion, the two differ. Diffusion grows with the capability gap, so faster diffusion closes part of a wider gap. A compute gap instead slows the last stretch of the catch-up, after diffusion has faded and own research does the work. Proposition~\ref{prop:buy} gives the two sensitivities in the constant-speed case.

Panel 2 makes the two countries equal in capability as well as compute, with both frontiers paused at the Automated Coder. The window \eqref{eq:game_physical_window} ties the strategic position to which country is ahead, so between equals it is zero and another measure is needed. Integrate the progress law as for \eqref{eq:catch_no_copy}, over the stretch the full-speed path covers in the first $\Delta$ months of the takeoff period and with $\kappa=1$. This gives \eqref{eq:tau}, and setting $\tau_\Delta(d)$ equal to the critical window $\xi/m_{\max}$ gives \eqref{eq:symmetric_date}. Matching the main example's date takes $\Delta=\symLeadStar$ months. With $\Delta=18$, a breakout is decisive only when the secret program completes the takeoff period, and cooperation is supportable until month \symLeadEighteenLast.

The game gives each country half of control in a contested race, $\chi_A=\chi_B=1/2$. A leader with a head start would likely win more. A larger share $\chi_i$ raises country $i$'s contested payoff $\mathcal R_i$ and its cooperative payoff $\mathcal C_i$ and leaves its breakout payoff $\mathcal T_i$ unchanged. So its breakout temptation $\xi_i$ falls and its gain from cooperating, $\mathcal C_i-\mathcal R_i$, rises. Panel 3 gives the leader between 45 and 60 percent. At 55 and 60 percent, the follower's temptation rises from \symStakesBaseXiC{} to \symStakesFiftyFiveXiC{} and \symStakesSixtyXiC. With its smaller gain from cooperating, the follower then sets both bounds. The critical window rises from \symStakesBaseCrit{} to \symStakesFiftyFiveCrit{} and \symStakesSixtyCrit{} months, and the last supportable review falls from month \symStakesBaseLast{} to \symStakesFiftyFiveLast{} and \symStakesSixtyLast. Unequal stakes put the temptation on the country with the least to gain from cooperating. Which way they move the date depends on whose window binds. After the start of the pause the follower's window is the shorter, so a larger share for the follower helps: at $\chi_A=0.45$ the follower needs only \symStakesFortyFiveCrit{} months and cooperation is supportable until month \symStakesFortyFiveLast. Between countries that differ only in their shares and have equal windows, equal shares minimize the critical window.

The main example fixes the leader's window at 12 months and derives only the follower's. A secret leader program also gains from growing compute. Read the 12 months as a distance, the progress of the first 12 months of the full-speed takeoff period. The leader's window is then $\tau_{12}(d)$ from \eqref{eq:tau}. It starts at 12 months, like the follower's, but falls faster and is the shorter of the two at every later review date. It then sets the clock, and the last supportable review moves from month \symLeaderFixedLast{} to \symLeaderTwelveLast{} (Panel 4), the same date as for equal countries with $\Delta=12$. The main example's date holds if a leader breakout must cover at least \symLeadStar{} months of full-speed progress before the follower loses its contested position. If the breakout is decisive only when it completes the takeoff period ($\Delta=18$), the follower's window binds first and the date stays at month \symLeaderEighteenLast. So the clock does not depend on the leader's window, but its date does.

In every row with a positive window, the binding window is shorter at month 24 than at the start, so the clock survives each of these changes. What moves is the date. A lead buys time to respond, because a follower must first erase it, and a follower that relies little on diffusion erases it faster. Stakes that favor the leader raise the time needed. If a wider lead also tilted the contested race toward the leader, these two effects would pull against each other. The model does not link them, so the table reports them separately.

\section{Notation}\label{app:notation}
\begin{table}[htbp]
\centering\small
\caption{Notation, in order of appearance}\label{tab:notation}
\begin{tabular}{@{}p{.27\textwidth}p{.68\textwidth}@{}}
\toprule
Symbol & Meaning\\
\midrule
$K(t),g_K,\gamma$ & Installed compute; its growth rate; growth of research capacity, $\varepsilon g_K$\\
$i,x,s$ & Shares for customer inference, capability research, and safety research\\
$F,\phi(F),\varepsilon$ & Capability (0 Automated Coder, 1 TED-AI); research productivity of a model of capability $F$; research elasticity\\
$S,\sigma(F),y(s)$ & Safety knowledge; safety yield; learning multiplier\\
$\theta,\rho,\psi,\alpha$ & Calibration forms $\phi(F)=\theta e^{\rho F}$ and $y(s)=\psi+(1-\psi)(s/s_0)^\alpha$\\
$h(F),k(F),\lambda(t),H,P$ & State and step hazard; hazard rate; cumulative hazard; catastrophe probability\\
$F_B,a_B,\gamma_B,\delta$ & Follower capability; its own research level and growth; share of the gap closed by diffusion per month\\
$G,G_0,G^*,T^*,u$ & Capability gap; its initial value; stationary gap; catch-up time after a halt; diffusion-to-research ratio $\delta G_0/a$\\
$\nu$ & Leader's constant capability pace\\
$\sigma_B$ & Follower safety yield; $\sigma/2$ in the baseline\\
$D,\ell,R$ & Detection delay; implementation delay; remaining agreed pause\\
$d,T_{\mathrm{catch}}(d),w_B(d)$ & Review date; catch-up time after a secret advance at $d$; the follower's response window\\
$i\in\{A,B\}$ & Country index: leader $A$, follower $B$\\
$\mathcal C_i,\mathcal T_i,\mathcal R_i,\mathcal S_i$ & Payoffs: cooperation, breakout, contested race, waiting while the rival advances\\
$Y_i,Z_i,\chi_i$ & Survival value; value of strategic control; share of control in a contested outcome\\
$c_i,d_i,p_0,p_1,p_2$ & Costs of development and of delay; catastrophe probability when 0, 1 or 2 countries advance\\
$m,\eta_i,w_i$ & Monitoring intensity (findings per month); cost per unit; response window\\
$\xi_i$ & Breakout temptation $\ln[(\mathcal T_i-\mathcal R_i)/(\mathcal C_i-\mathcal R_i)]$\\
$m_{\min},m_{\max}$ & Least intensity that deters both countries; most both will pay\\
$\Delta,\tau_\Delta(d),d^*_\Delta$ & Between equals: the lead that makes a breakout decisive (months of full-speed progress); the window; the last supportable review date\\
$\Lambda_j,I_j(T)$ & Cumulative hazard of frontier $j$; cause-specific catastrophe incidence from frontier $j$\\
$r$ & Monthly discount rate\\
$L,\kappa$ & Appendix~\ref{app:symmetric}: the follower's lag in months of full-speed progress; its research capacity relative to the leader's\\
\bottomrule
\end{tabular}
\end{table}

\FloatBarrier
\section{Proofs}\label{app:proofs}

\subsection{The Follower and the Lead}

\subsubsection{Proof of Proposition~\ref{prop:copy}}\label{app:proof_copy}

\begin{proof}
With $\dot F = 0$, the gap $G(t) = F - F_B(t)$ satisfies $\dot G = -\dot F_B = -a - \delta G$. This linear ODE has solution
\begin{equation*}
G(t) = \left(G_0 + \frac{a}{\delta}\right)e^{-\delta t} - \frac{a}{\delta}.
\end{equation*}
Setting $G(T^*) = 0$:
\begin{equation*}
e^{-\delta T^*} = \frac{a/\delta}{G_0 + a/\delta} = \frac{1}{1 + \delta G_0 / a} = \frac{1}{1 + u} \implies T^* = \frac{1}{\delta}\ln(1 + u),
\end{equation*}
which proves \eqref{eq:catchup_formula}. When the follower keeps pace, its speed is $a + \delta G_0$. The raw lead (months of progress separating the two) is $\text{raw} = G_0/(a + \delta G_0) = G_0/[a(1 + u)]$. Dividing $T^*$ by $\text{raw}$:
\begin{equation*}
\frac{T^*}{\text{raw}} = \frac{\frac{1}{\delta}\ln(1+u)}{\frac{G_0}{a(1+u)}} = \frac{a(1+u)\ln(1+u)}{\delta G_0} = \frac{(1+u)\ln(1+u)}{u},
\end{equation*}
which establishes \eqref{eq:ratio_formula}.
\end{proof}

\subsubsection{Proof of Proposition~\ref{prop:overtake}}\label{app:proof_overtake}

\begin{proof}
The gap evolves according to $\dot G = \nu - \dot F_B = \nu - a - \delta G$. The stationary point is $G^* = (\nu - a)/\delta$. The general solution is $G(t) = G^* + (G_0 - G^*)e^{-\delta t}$, proving (a).

If $\nu>a$, then $G^*>0$ and $G(t)\ge\min\{G_0,G^*\}>0$. If $\nu=a$, the solution is $G(t)=G_0e^{-\delta t}>0$ at every finite date. If $\nu<a$, the pre-crossing solution reaches zero at $t_{\text{cross}}=\delta^{-1}\ln[(G_0-G^*)/(-G^*)]$. Beyond that date the diffusion term $\delta\max\{G,0\}$ is zero and the linear solution in (a) no longer applies. Part (c) follows from applying \eqref{eq:catchup_formula} while the gap is positive.
\end{proof}

\subsection{The Response Window and the Agreement Game}

\subsubsection{Lemma~\ref{prop:mutual} and Its Proof}\label{app:proof_mutual}

\begin{lemma}[Detection and Lead-Preserving Restart]\label{prop:mutual}
In the constant-speed environment of Proposition~\ref{prop:copy}, let the gap when the follower secretly advances be $G_0>0$. Let $D$ be detection delay, $\ell\ge0$ the implementation delay, and $R$ the remaining agreed pause. Suppose the leader restarts at the earlier of the completed alarm response and the scheduled pause end, so its actual unilateral exposure is $\min\{D+\ell,R\}$.
\begin{enumerate}
\item[(a)] The leader has a positive frontier gap at the restart if and only if $\min\{D+\ell,R\}<T^*(G_0)$.
\item[(b)] If this holds and the leader then maintains speed $\nu\ge a$, the follower does not overtake during the remaining development path.
\end{enumerate}
This is a physical resilience result; it does not establish either country's preference for compliance.
\end{lemma}

\begin{proof}
Before the restart, the gap follows the unilateral-halt law in Proposition~\ref{prop:copy} and first reaches zero at $T^*(G_0)$. So the gap at the restart is positive exactly when $\min\{D+\ell,R\}<T^*(G_0)$. Restarting from that positive gap at $\nu\ge a$ preserves positivity by Proposition~\ref{prop:overtake}.
\end{proof}

\subsubsection{Proof of Proposition~\ref{prop:credible_response}}\label{app:proof_credible_response}

\begin{proof}
The monitoring payment is sunk. Restarting is preferred exactly when $\mathcal R_j>\mathcal S_j$. Substituting \eqref{eq:game_payoffs} gives \eqref{eq:credible_response}.
\end{proof}

\subsubsection{Proof of Proposition~\ref{prop:small_agreement}}\label{app:proof_small_agreement}

\begin{proof}
Backward induction gives restarting after a timely finding by Proposition~\ref{prop:credible_response}. If the rival complies, unilateral advancement pays
\[
e^{-mw_i}\mathcal T_i+(1-e^{-mw_i})\mathcal R_i-\eta_i m,
\]
whereas compliance pays $\mathcal C_i-\eta_i m$. Compliance is optimal exactly when $e^{-mw_i}(\mathcal T_i-\mathcal R_i)\le\mathcal C_i-\mathcal R_i$. Taking logs, this is $mw_i\ge\ln[(\mathcal T_i-\mathcal R_i)/(\mathcal C_i-\mathcal R_i)]=\xi_i$, or $m\ge\xi_i/w_i$. Both conditions give $m_{\min}$. Entry is preferable to the no-agreement race exactly when $\mathcal C_i-\eta_i m\ge\mathcal R_i$ for both countries, giving $m_{\max}$. So the feasible interval is necessary and sufficient.

For existence in the proposal game, select mutual compliance after acceptance when $m$ lies in this interval. Outside it select mutual advancement in the action subgame; this is always an equilibrium, because pausing against advancement pays $(1-e^{-mw_j})\mathcal R_i+e^{-mw_j}\mathcal S_i-\eta_i m<\mathcal R_i-\eta_i m$ at finite $m$. The follower accepts a feasible proposal and rejects positive-cost proposals with the racing continuation. The leader prefers the lowest feasible $m$, since its cooperative payoff decreases with $m$, and prefers that proposal to refusal. Tie breaking supports the weak boundary. An intensity strictly inside the interval makes both compliance and participation strict. The action game also permits a racing continuation at feasible proposals; selecting it can prevent agreement.
\end{proof}

\subsubsection{Proof of Corollary~\ref{cor:game_window}}\label{app:proof_game_window}

\begin{proof}
Equation~\eqref{eq:game_interval} requires $\xi_B/w_B(d)\le m_{\max}$, giving \eqref{eq:game_window_bound}. After a secret advance at date $d$, the follower's capability $v$ months later follows
\[
\frac{dF_B}{dv}=a_Be^{\gamma_B(d+v)}\frac{\phi(F_B)}{\phi(0)}
+\delta(F-F_B)
\]
until catch-up. Starting from the same fixed $F_B(0)$ and $F$, a later $d$ strictly increases own progress at every pre-crossing state. So scalar comparison gives an earlier crossing. Let $\phi_*>0$ be the minimum of $\phi$ on $[F_B(0),F]$. Since progress before crossing is at least $a_Be^{\gamma_B d}\phi_*/\phi(0)$, the initial gap $G_0$ gives
\[
0<T_{\mathrm{catch}}(d)\le
\frac{G_0\,\phi(0)}{a_Be^{\gamma_B d}\phi_*}\longrightarrow0.
\]
The positive right side of \eqref{eq:game_window_bound} then excludes dates that are late enough.
\end{proof}

\subsection{Widening the Lead}

\subsubsection{Proof of Proposition~\ref{prop:buy}}\label{app:proof_buy}

\begin{proof}
Differentiating \eqref{eq:catchup_formula} in $a$:
\begin{equation*}
\frac{\partial T^*}{\partial a} = \frac{1}{\delta}\,\frac{1}{1 + \delta G_0 / a}\,\left(-\frac{\delta G_0}{a^2}\right) = -\frac{G_0}{a(a + \delta G_0)}.
\end{equation*}
Multiplying by $a$ yields $\frac{\partial T^*}{\partial \ln a} = -\frac{G_0}{a + \delta G_0} = -\text{raw lead}$.

Next, differentiating in $\delta$:
\begin{equation*}
\frac{\partial T^*}{\partial \delta} = -\frac{1}{\delta^2}\ln(1 + u) + \frac{1}{\delta}\,\frac{G_0/a}{1 + u} = -\frac{T^*}{\delta} + \frac{G_0}{a + \delta G_0}\,\frac{1}{\delta}.
\end{equation*}
Multiplying by $\delta$ yields $\frac{\partial T^*}{\partial \ln \delta} = -T^* + \text{raw lead} = -(T^* - \text{raw lead})$.
\end{proof}

\end{document}